\documentclass[a4paper,11pt]{article}

	\usepackage{enumerate,hyperref,siunitx}
	\usepackage{enumitem,}
	\usepackage{siunitx}
	\usepackage{cite}
	\usepackage{bm,amsmath,amssymb,physics}
	\usepackage{graphicx}
	\usepackage[font=small,labelfont=bf,figurename=Figure,labelsep=period]{caption}
	\usepackage{multirow,tabularx}
	\usepackage[margin=1in]{geometry}
	\usepackage{float,pdflscape,authblk,setspace,lineno}	
	\usepackage{xcolor}
	\usepackage[symbol]{footmisc}
	\newcommand{\killpunct}[1]{}
	
	\usepackage[capitalise]{cleveref}
	\usepackage{placeins}

\begin{document}

	\title{Predicting radiotherapy patient outcomes with real-time clinical data using mathematical modelling}


	\author[1]{Alexander P. Browning\textsuperscript{*}\textsuperscript{$\ddagger$}}
	\author[1,2]{Thomas D. Lewin\textsuperscript{*}}
	\author[1]{Ruth E. Baker}
	\author[1]{Philip K. Maini}
	\author[3]{Eduardo G. Moros}
	\author[3]{Jimmy Caudell}
	\author[1]{Helen M. Byrne\textsuperscript{$\dagger$}}
	\author[3,4,5]{Heiko Enderling\textsuperscript{$\dagger$}}

	\affil[1]{Mathematical Institute, University of Oxford, Oxford, UK}
	\affil[2]{Roche Pharma Research and Early Development, Roche Innovation Center, Basel, Switzerland}
	\affil[3]{Department of Radiation Oncology, H. Lee Moffitt Cancer Center \& Research Institute, USA}
	\affil[4]{Department of Integrated Mathematical Oncology, H. Lee Moffitt Cancer Center \& Research Institute, USA}
	\affil[5]{New Address: Department of Radiation Oncology, MD Anderson Cancer Center, Houston, TX, USA}


\date{\today}
\maketitle
\footnotetext[1]{These authors contributed equally.}
\footnotetext[2]{These authors also contributed equally.}
\footnotetext[3]{Corresponding author: browning@maths.ox.ac.uk or HEnderling@mdanderson.org}

	\vfill
	\renewcommand{\abstractname}{Abstract}
	\begin{abstract}
		\noindent
		Longitudinal tumour volume data from head-and-neck cancer patients show that tumours of comparable pre-treatment size and stage may respond very differently to the same radiotherapy fractionation protocol. Mathematical models are often proposed to predict treatment outcome in this context, and have the potential to guide clinical decision-making and inform personalised fractionation protocols. Hindering effective use of models in this context is the sparsity of clinical measurements juxtaposed with the model complexity required to produce the full range of possible patient responses. In this work, we present a compartment model of tumour volume and tumour composition, which, despite relative simplicity, is capable of producing a wide range of patient responses. We then develop novel statistical methodology and leverage a cohort of existing clinical data to produce a predictive model of both tumour volume progression and the associated level of uncertainty that evolves throughout a patient's course of treatment. To capture inter-patient variability, all model parameters are patient specific, with a bootstrap particle filter-like Bayesian approach developed to model a set of training data as prior knowledge. We validate our approach against a subset of unseen data, and demonstrate both the predictive ability of our trained model and its limitations.
	\end{abstract}
	\vfill

	\renewcommand{\abstractname}{Keywords}
	\begin{abstract}
		\noindent 
		\centering
		head-and-neck cancer, predictive model, patient variability, heterogeneity, uncertainty, radiotherapy
	\end{abstract}
	\vspace{1cm}
	\vfill

\section{Introduction}

	Radiotherapy remains a mainstay of cancer treatment, with approximately half of all cancer patients receiving radiotherapy as part of their standard of care \cite{Fowler.2006,Torres-Roca.2012,Enderling.2009}. It is common for a patient's course of treatment to be determined solely by tumour etiology, location, and stage. Other patient-specific factors, such as the intrinsic radiosensitivity and composition of a tumour, are not typically used to inform protocol selection in the clinic \cite{Caudell.2017}. Clinical studies suggest that patients at a similar tumour, node, and metastasis (TNM) stage, and with comparable pre-treatment tumour volumes, may respond differently to the same radiotherapy fractionation schedule \cite{Scott.2017,Sunassee.2019}. Mathematical models have the potential to capitalise on real-time clinical observations to both predict patient specific responses and guide clinical decision-making. It is hoped that such a tight integration could eventually be used to personalise fractionation schedules either \textit{a priori} or adaptively during a patient's course of treatment \cite{Enderling.2019}.
	
	Challenges associated with the application of mathematical models to interpret data and draw predictions are perhaps most acute for single-patient clinical data. Models must be sufficiently complex to reproduce the full gamut of patient responses \cite{Yankeelov.2013,Collis.2017,Brady.2019}. However, clinical data are often limited, typically comprising solely noisy measurements of the gross tumour volume (GTV) at sparse time intervals throughout a patient's course of treatment \cite{Brady.2019,Harshe.2023}. The necessity to start treatment as soon as possible after diagnosis means that pre-treatment predictions are often drawn from only one or two observations. Consequently, models aimed at clinical applications are relatively simple \cite{Prokopiou.2015,Rockne.2017iu}, incorporate limited biological detail, and often describe only the time-evolution of the GTV \cite{Sunassee.2019,Prokopiou.2015,Rockne.2017iu}. While simplicity can elicit parameter identifiability and avoid overfitting, predictions can be poor---or even misleading---if a model is so simple as to be unable to capture the range of observed (possible) responses. The dangers of overfitting are particularly pronounced for single-patient clinical data used for prediction, where model validation must be assessed pre-treatment; in diametric opposition to experimental data, technical replicates are never available. It is, therefore, crucial to validate models across a wide range of responses, and to accurately quantify uncertainty in predictions used in clinical decision-making \cite{Brady.2019}.

	\begin{figure}[!t]
		\centering
		\includegraphics[width=\textwidth]{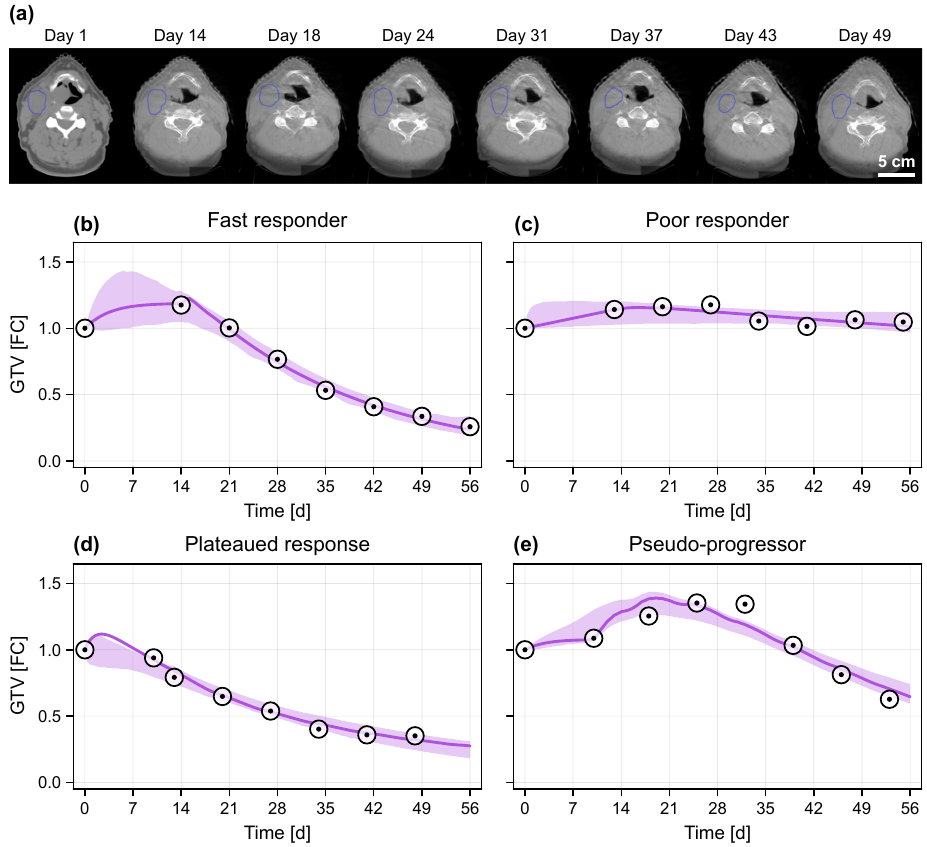}
		\caption[Figure 1]{\textbf{Gross tumour volume (GTV) measurements taken during radiotherapy.} (a) Example CT scan of an oropharyngeal cancer patient throughout treatment, showing tumour contoured in blue (data courtesy of CD Fuller and MD Anderson). (b--e) Clinical data representing four qualitatively different radiotherapy results. Predictions from the mathematical model, along with a 95\% credible interval for the modelled observation means, are shown in purple. In all cases, the radiotherapy schedule starts at the time of the second observation. The four patients shown are excluded from the training data analysed in later parts of our study. The units of GTV are given as the fold change (FC) relative to the initial volume.}
		\label{fig1}
	\end{figure}

	In this work, we present a predictive mathematical modelling framework using clinical GTV data from a previously published cohort of head-and-neck cancer patients who exhibit a variety of treatment responses (\cref{fig1}) \cite{Zahid.2021nem,Zahid.2021}. The primary goal of our framework is to integrate previously observed clinical observations to predict the time course of radiotherapy response in new patients. To demonstrate our framework, we focus our analysis on prediction of the tumour volume progression in four patients presented in \cref{fig1} and in our previous work \cite{Lewin.2020}: these patients are excluded from the otherwise randomly-selected cohort of patients used to train the mathematical model. All patients in the clinical data set receive a standard radiotherapy fractionation schedule, comprising fractions of \SI{2}{\gray} delivered on weekdays over a four- to seven-week period \cite{Lewin.2016}. To keep our study as widely applicable as possible, we work with the most fundamental, albeit limited, mode of single patient data. Computed tomography (CT) scans are routinely used to image tumours pre-treatment at both the diagnosis and treatment planning stages (\cref{fig1}a)  \cite{Stevens.2013,Sharma.2016,Wang.2009}. Further scans, such as cone beam CT, may be taken upon the delivery of each fraction but are usually used solely for alignment purposes; for our data, scans were available once per week during treatment. These CT scans are not of a high spatial resolution, are noisy, of a low contrast, and do not differentiate heterogeneity in tumour composition. As such, only noisy measurements related to an estimate of the GTV are available at relatively sparse intervals throughout each patient's course of treatment (typically, once per week). The heterogeneity in radiotherapy response exhibited in \cref{fig1}b--e raises several important questions: in particular, how early into treatment can a practitioner determine if a patient is responsive, and to what extent is it possible to predict the final tumour volume during treatment using only GTV measurements?  Given the side-effects associated with radiotherapy, and possible indirect costs of switching treatments at too late a TNM stage, any improvement in prediction accuracy is of great clinical value.

	Mathematical models of tumour progression vary significantly in complexity; ranging from simple phenomenological models of GTV, such as logistic and Gompertz growth \cite{Sachs.2001,McAneney.2007,Rockne.2010,Chvetsov.2013,Prokopiou.2015,Tariq.2016,Poleszczuk.2018,Browning.2023}, to highly detailed spatial models that capture multiple facets of tumour heterogeneity \cite{Greenspan:1972oq,Rockne.2010,Lewin.2018,Lewin.2020, Browning.2023}. The limitations and challenges imposed by clinical data yield an overrepresentation of the former, meaning that the functional forms for both growth and radiotherapy response are motivated almost entirely by empirical observations rather than the underlying biological mechanisms. Yet, it is now well established that intra-tumour heterogeneity and the tumour microenvironment play important roles in overall growth, and may significantly influence treatment outcome \cite{Ribba.2006,Rockne.2009,Rockne.2015,Lewin.2018,Lewin.2020,Browning.2021eqr}. Motivated by these findings and in consideration of the noisy data available for prediction, we take an intermediate approach and utilise a two compartment extension of the so-called proliferation-saturation-index (PSI) model of Prokopiou et al. \cite{Prokopiou.2015} and later Poleszczuk et al. \cite{Poleszczuk.2018}. This choice of ordinary differential equation (ODE) model balances simplicity, through a phenomenological description of radiation-free tumour growth saturation, with biological detail, through a radiotherapy response corresponding to a transfer of cellular material from a living to a dead state. Compared with purely statistical or machine learning models, our mathematical approach allows a full interpretable integration of clinical data from individuals, whereby the radiotherapy schedule is imported directly from the reported patient fractionation schedule. Finally, our model contains sufficient detail to allow us to quantify the potential utility of expanding clinical data collection to include information relating to tumour composition in addition to GTV.

\begin{figure}
		\centering
		\includegraphics{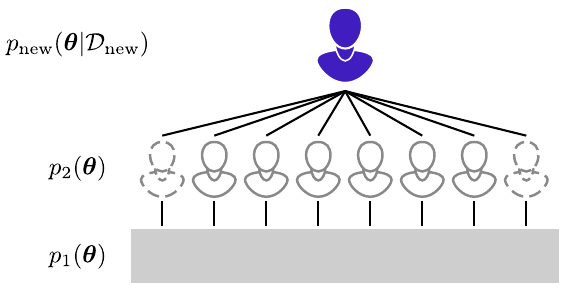}
		\caption[Figure 2]{\textbf{Pseudo-hierarchical approach used in the analysis.} Devoid of any data, knowledge about model parameters is encoded in the ``first-level prior'', denoted by $p_1(\bm\theta)$, and is used to individually form a set of posterior distributions for patients in the training set. The ``second-level prior'', denoted by $p_2(\bm\theta)$, represents knowledge gained from analysis of the training set and is used to form posterior distributions for \textit{new} patients, denoted by $p_\text{new}(\bm\theta|\mathcal{D}_\text{new})$. In effect, the approach identifies patients in the training set with possibly similar outcomes to new patients.}
		\label{fig2}
	\end{figure}

	We take a Bayesian pseudo-hierarchical approach to inference and model calibration, by leveraging observed population-level information to draw predictions and quantify corresponding levels of prediction uncertainty. To account for inter-patient heterogeneity, all model parameters are allowed to vary between patients. A schematic of the approach is provided in \cref{fig2}. In contrast to standard Bayesian hierarchical approaches, we do not make any parametric assumptions relating to the distribution of model parameters between individuals. Instead, we build up a population-level posterior distribution by calibrating the model to individual patients in a set of training data. As illustrated in \cref{fig2}, this population-level posterior distribution forms the prior for analysis of new patients. In effect, the prior distribution for new patients quantifies a set of possible patient outcomes based on those observed in the training data. Successively applying the Bayesian inference algorithm as data becomes available throughout a patient's course of treatment allows us to update this potential set of future outcomes and the corresponding uncertainty in tumour volume. We validate our approach by first exploring prediction ability on synthetic data, and then prospectively making predictions on the four patients presented in \cref{fig1}b--e as they undergo their course of treatment.

\section{Methods}

In this section, we outline the clinical data, and the mathematical and statistical methodology developed and later employed in this work. First, in \cref{methods_exp} we describe and present the clinical data set used for quantitative analysis and which demonstrate four disparate treatment response classifications. Secondly, in \cref{methods_maths} we present a mechanistic mathematical model of tumour volume progression, along with a set of objective criteria that we use to classify model realisations into the four observed classifications. Subsequently, in \cref{methods_stats} we present a statistical model that connects model predictions to clinical measurements. In \cref{methods_inference} we outline the novel statistical methodology employed in the analysis. Finally, in \cref{synthetic_data_methods} we outline the procedure for resampling from the joint posterior to produce synthetic patient data. A Julia implementation of the model and inference algorithm, along with data used in the analysis, are available on GitHub\footnote{\url{https://github.com/ap-browning/clinical-inference}}.

\subsection{Tumour volume data}\label{methods_exp}

Current clinical practice involves two CT scans collected for each patient; one at diagnosis and one at treatment planning. These scans are then used to estimate GTV \cite{Wang.2009,Stevens.2013,Sharma.2016}. While it is feasible to obtain further scans at the time of delivery of each fraction, these scans are often of a low quality, being used primarily to position the patient. As such, they are not typically stored for research purposes.

In this paper, we use retrospective volumetric data, collected weekly, from head-and-neck cancer patients, across multiple anatomical locations, including the oropharynx, tonsil and base of tongue. Patients were immobilised via a thermoplastic mask with or without bite block. Isocenter and positioning was verified daily via orthoganal kV or CBCT imaging. Each CT scan was segmented by the same radiation oncologist, giving weekly tumour volumes throughout treatment in addition to a volume measurement at the treatment planning stage. Weekly cone beam CT (CBCT) scans were extracted from the record and verify system (Mosaiq, Elekta).  Suitable CBCTs with minimal artifact were selected for contouring. Clinical target volume (CTV) was created from GTV with a \SI{5}{\milli\metre} isotropic expansion.  CTV was then trimmed from barriers to spread including air, bone, fascial planes, and in some cases muscle.  Planning target volume (PTV) was created from CTV via \SI{3}{\milli\metre} isotropic expansion. An example suite of contoured CT scans from a single patient is shown in \cref{fig1}a. The GTV data shown in \cref{fig1}b--e correspond to those presented and discussed in Lewin et al. \cite{Lewin.2020}. In total, GTV data from 51 patients was collected and made available as supplementary material. All methods were carried out in accordance with the institutional policies of the Moffitt Cancer Center. The clinical protocol covering patient data and methods used in this paper was approved by the Moffitt Cancer Center's Institutional Review Board (IRB). Since this is a retrospective study using de-identified data of adult human subjects, informed consent was waived by the IRB.

\subsection{Mathematical model}\label{methods_maths}

Mathematical models of tumour growth and, to a lesser extent, radiotherapy response, are well established \cite{Araujo.2003auos}, ranging from compartmental ODE models \cite{Sachs.2001,Chvetsov.2009,Wang.2013,Chvetsov.2014,Prokopiou.2015,Tariq.2016,Sunassee.2019,Browning.2023} and spatially-resolved partial differential equation models \cite{Greenspan:1972oq,Rockne.2009,Rockne.2010,Rockne.2015,Lewin.2018,Browning.2021eqr,Browning.2023}, to agent-based models \cite{Enderling.2009,Gao.2013,Alfonso.2014,Powathil.2013,Powathil.2016,Richard.2007} and purely probabilistic models \cite{Zaider.2000,Hanin.2004,Gong.2011,Bobadilla.2017}.

Given the limitations imposed by GTV clinical data, we present a relatively simple mathematical model that is able to capture the four classes of tumour response observed in \cref{fig1}. In particular, we extend the PSI model \cite{Poleszczuk.2018} to include a simple measure of tumour composition by modelling the volume of both living cells, $L(t)$, and necrotic debris, $N(t)$. The GTV is given by $V(t) = L(t) + N(t)$. Living cells proliferate logistically with rate $\lambda\,\SI{}{\per\day}$ and carrying capacity $K$ $[V(t)]$, and potentially undergo necrosis at rate $\eta\,\SI{}{\per\day}$. Given that the growth dynamics occur on a much slower timescale than the interval during which the patient receives each fraction, we model radiotherapy as an instananeous transfer of living cells to necrotic debris at record-informed dosing times $t_i$, $i = 1,2,\dots,n$. The model equations are given by
	\begin{equation}
	\begin{aligned}
		\dv{L}{t} &= \overbrace{\vphantom{\Bigg(}\lambda L\left(1 - \dfrac{L}{K}\right)}^{\text{Growth}}\hspace{3mm} - \overbrace{\vphantom{\Bigg(}\eta L}^{\text{Necrosis}} -\hspace{2mm} \overbrace{\vphantom{\Bigg(}\gamma L \sum_{i=1}^n \delta(t - t_i)}^{\text{Radiotherapy}},\\
		\dv{N}{t} &= \eta L \hspace{1mm}- \underbrace{\vphantom{\Big(}\zeta N}_{\text{Decay}} + \hspace{1mm}\gamma L \sum_{i=1}^n \delta(t - t_i),\\
	\end{aligned}
	\end{equation}
where $\delta(t - t_i)$ is a delta function, representing a transfer of a volume $\gamma L$ from the living compartment to the dead compartment, such that $\gamma\,\SI{}{\per\day}$ quantifies the strength of radiotherapy response. We assume further that necrotic material is degraded at a constant rate $\zeta\,\SI{}{\per\day}$. To capture inter-patient heterogeneity, all parameters are allowed to vary between patients \cite{Lawson.2018}.

The data suggest that initial GTV is comparable between responsive and poorly responsive patients (\cref{tab1}). Therefore, we normalise $L(t)$ and $N(t)$ with the initial GTV such that $V(0) = 1$ and describe the initial tumour composition as 
	\begin{equation}
		L(0) = 1 - \phi_0,\qquad N(0) = \phi_0, 
	\end{equation}
where $0 \le \phi_0 \le 1$ is an unknown, patient-specific parameter to be estimated that represents the proportion of the tumour occupied by dead material at $t = 0$. We note further that the interpretation of the carrying capacity parameter $K$ is with respect to the measured initial GTV. Thus, GTV measurements presented throughout the paper may be interpreted as the fold change (FC) compared to the initial GTV. The interpretation of all other parameters remains unchanged by this choice of units.

In the supplementary material (Figs. S1 and S2), we perform a parameter sweep across parameters relating to necrosis and necrotic material decay ($\eta$ and $\zeta$, respectively), for a patient subject to daily doses of radiotherapy on weekdays over a six week period, to verify that the model is able to reproduce the wide range of dynamics observed in the clinical data. While the parameter sweep is not exhaustive, the results demonstrate that varying only these two parameters is sufficient to produce the range of responses observed in \cref{fig1}.

	\begin{table}[!b]
		\centering
		\caption[Table 1]{Prior classification of each patient response class, based on the full posterior, $p(\bm\theta | \{\mathcal{D}_i\}_{i=1}^n)$ and the second-level prior $p_2(\bm\theta)$, the latter corresponding to an expanded kernel density estimate constructed from samples of the full posterior. The statistics related to the initial volume are based on the classifications of the prior samples corresponding to each patient in the training set, hence non-integer counts arise due to probabilistic classification of patients. An approximate statistical test, based on Welch's approximate unequal variance $t$-test \cite{Welch.1947oy}, indicates no statistically significant difference between fast and poor responders ($P = 0.582$), nor between responders ($*$) and poor responders ($P = 0.557$). Asterisks indicate classifications corresponding to patients who show an eventual response.}
		\label{tab1}
		\begin{tabular}{lccccccc}\hline
			\multirow{2}{*}{Classification} & \multicolumn{2}{c}{Proportion} & \multicolumn{5}{c}{Initial Volume [cm\textsuperscript{3}]}\\
			 & $p(\bm\theta | \{\mathcal{D}_i\}_{i=1}^n)$ & $p_2(\bm\theta)$ &\hphantom{M}& Mean & Std & Count\\\hline\hline
			($\ast$) Fast responder 		& 0.8763 & 0.6278 && 16.8 & 11.6 & 35.0&\\
			Poor responder 					& 0.0598 & 0.3473 && 20.2 & 7.3  & 2.4&\\
			($\ast$) Plateaued response 	& 0.0035 & 0.0021 && 3.4  & 4.8  & 0.1&\\
			($\ast$) Pseudo-progression 	& 0.0604 & 0.0228 && 13.7 & 12.0 & 2.4&\\\hline
			Eventual response ($\ast$) 		& 0.9402 & 0.6527 && 16.6 & 11.7 & 37.6\\\hline
		\end{tabular}
	\end{table}

\subsubsection{Classifying responses}\label{sec:classification}

We observe four classes of qualitative response within the clinical data, as highlighted in \cref{fig1} and summarised in \cref{tab1}. In \cref{fig1}b, the patient responds well to radiotherapy, with the tumour decreasing markedly in volume throughout treatment. Hereafter, we refer to a patient exhibiting this type of behaviour as a \textit{fast responder}. By contrast, there are patients for whom the effects of radiotherapy appear to be marginal when viewed in terms of tumour volume over time alone, as is the case in \cref{fig1}c. We classify these patients as \textit{poor responders}. In a number of cases, the initial response of the tumour to radiotherapy appears to be favourable, but the response plateaus in the latter stages of treatment, resulting in a non-negligible final tumour volume (\cref{fig1}d). Such patients are classified as having a \textit{plateaued response}. However, this radiographic volume may subsequently recede in the weeks after radiotherapy. Occasionally, as in \cref{fig1}e, a patient may appear to exhibit continued tumour progression throughout the first few weeks of radiotherapy before showing a delayed response, characterised by a decrease in tumour volume towards the end of treatment. We characterise this type of response as \textit{pseudo-progression}. 

We classify a model realisation into one of four classes of response based on a standard patient receiving doses on weekdays over a six week period, with CT measurements taken at the start of each treatment week and at the time of the final dose (the pre-treatment volume measurement is not used to classify patients). Based on the set of noise free synthetic measurements generated from the model, we define each classification according to the following quantitative criteria.

	\begin{enumerate}
		\item \textit{Poor responder.} All measurements above 85\% of the volume observed at the start of treatment.
		\item \textit{Responder.} At least one measurement below 85\% of the volume observed at the start of treatment. Responders are further classified:
		\begin{enumerate}
			\item \textit{Pseudo-progressor.} A second (noise-free) measurement greater than 102\% of the first following radiotherapy onset.
			\item \textit{Plateaued response.} Not a pseudo-progressor, with a final measurement greater than 20\% of the initial, and with a final rate-of-change less than 10\% of the maximum rate-of-change observed.
			\item \textit{Fast responder.} Not in any other classification. 
		\end{enumerate}
	\end{enumerate}

	The specific thresholds chosen in the classification algorithm yield excellent results that reliably distinguish between each class (Fig. S4). However, the relatively small number of plateaued responders and pseudo-progressors in the training set (\cref{tab1}) suggests that the criteria will need to be reassessed should more data become available.
		
\subsection{Statistical model}\label{methods_stats}

We take a standard approach and assume that CT scan data are independent and normally distributed about the model prediction \cite{Kreutz.2012} such that
	\begin{equation}
		V_\text{obs} \sim \mathcal{N}\left(V_\text{total}, \sigma^2(V_\text{total})\right),
	\end{equation}
where the standard deviation
	\begin{equation}
		\sigma(V_\text{total}) = \alpha_1 + \alpha_2 V_\text{total},
	\end{equation}
is assumed to be a linear function such that the statistical model captures both additive and multiplicative normal noise: $\alpha_1$ represents an absolute contribution to the variance, and $\alpha_2$ a relative contribution.

While the dynamical parameters are assumed to vary between patients, we assume that the noise parameters remain fixed. Therefore, we pre-estimate the noise parameters $\alpha_1$ and $\alpha_2$ by first inferring them alongside dynamical parameters for each patient. We then pool an equal number of noise parameter posterior samples for each patient and approximate $(\alpha_1,\alpha_2)$ as the marginal posterior mode. We are motivated to take this relatively standard approach of pre-estimating the noise parameters to reduce both the dimensionality of the parameter space and the complexity of the statistical methodology.

\subsection{Bayesian inference}\label{methods_inference}

An important difference between clinical and experimental data relates to the sample size: in clinical studies, each patient undergoes therapy only once. Given that patients are highly heterogeneous and data are relatively limited (\cref{fig1}), this poses a significant statistical challenge for computational inference. To account for this, we take a pseudo-hierarchical approach to inference and prediction by first training the model on a subset of the data (the training set). We are motivated to develop this novel approach to inference as opposed to a more standard Bayesian hierarchical approach as there is no sensible means by which to propose a particular distributional form for the joint parameter distributions at the population-level: given the distinct classes of response observed in \cref{fig1}, for example, we expect the joint parameter distribution to be multimodal. The correlation structure between model parameters is also unclear.

\begin{figure}
		\centering
		\includegraphics[width=\textwidth]{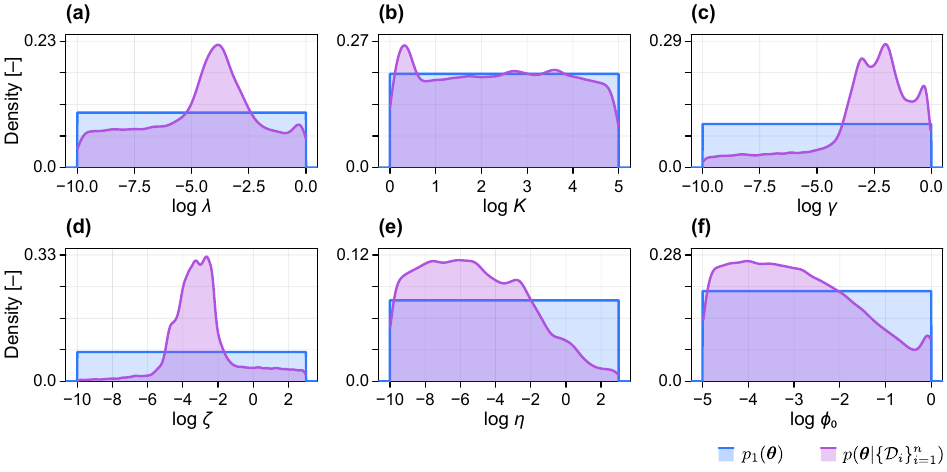}
		\caption[Figure 3]{\textbf{Parameter posteriors from analysis of training data.} first-level prior distribution (blue) and full posterior (purple) following analysis of the training data. The first-level prior, $p_1(\bm\theta)$, comprises independent uniform distributions in the log of each unknown parameter. Parameters relate to the cell proliferation rate, $\lambda$, the carrying capacity, $K$, the radiotherapy response strength, $\gamma$, the decay rate of necrotic debris, $\zeta$, the cell necrosis rate, $\eta$, and the initial proportion of the population that is necrotic, $\phi_0$.}
		\label{fig3}
	\end{figure}

From a full cohort of 51 patients, we randomly select a group of 40 patients to act as the \textit{training} set; these patients represent those that have been observed throughout an entire course of treatment, prior to the present. For each patient in the training set, we assume that initial knowledge about the model parameters is encoded in a ``first-level prior'', $p_1(\bm\theta)$, where $\bm\theta = (\log \lambda, \log K, \log \gamma, \log \zeta, \log \eta, \log \phi_0)$ (\cref{fig2}). We then update our knowledge about the parameters pertaining to patient $i$ using Bayes theorem such that
	\begin{equation}
		\underbrace{p^{(i)}(\bm\theta | \mathcal{D}_i)\vphantom{\Big(}}_{\text{Posterior $i$}} \propto \underbrace{p(\mathcal{D}_i | \bm\theta)\vphantom{\Big(}}_{\text{Likelihood}} p_1(\bm\theta),
	\end{equation}
where $\mathcal{D}_i$ represents data (including both volume measurements and the radiotherapy schedule) for patient $i$. We choose $p_1(\bm\theta)$ to an independent multivariate uniform (see \cref{tab2} and \cref{fig3}), an uninformative choice. 

	\begin{table}
		\centering
		\caption[Table 2]{Parameters and first-level prior distributions. The description relates to the exponentiated log parameter.}
		\label{tab2}
		\begin{tabular}{ccll}\hline
			Parameter 		& Units & \multicolumn{1}{c}{Prior} & Description \\\hline\hline
			$\lambda$ & \SI{}{\per\day} & $\log \lambda \sim \mathcal{U}(-10,0)$ 	& Cell proliferation rate\\
			$K$ & $-$ & $\log K \sim \mathcal{U}(0,5)$	& Carrying capacity\\
			$\gamma$ & \SI{}{\per\day} & $\log \gamma \sim \mathcal{U}(-10,0)$	& Radiotherapy response\\
			$\zeta$ & \SI{}{\per\day} & $\log \zeta \sim \mathcal{U}(-10,3)$	& Necrotic debris decay rate\\
			$\eta$ & \SI{}{\per\day} & $\log \eta \sim \mathcal{U}(-10,3)$	& Cell necrosis rate\\
			$\phi_0$ & $-$ & $\log \phi_0 \sim \mathcal{U}(-5,0)$	& Initial necrotic proportion\\\hline
		\end{tabular}
	\end{table}
	
\begin{figure}
		\centering
		\includegraphics[width=\textwidth]{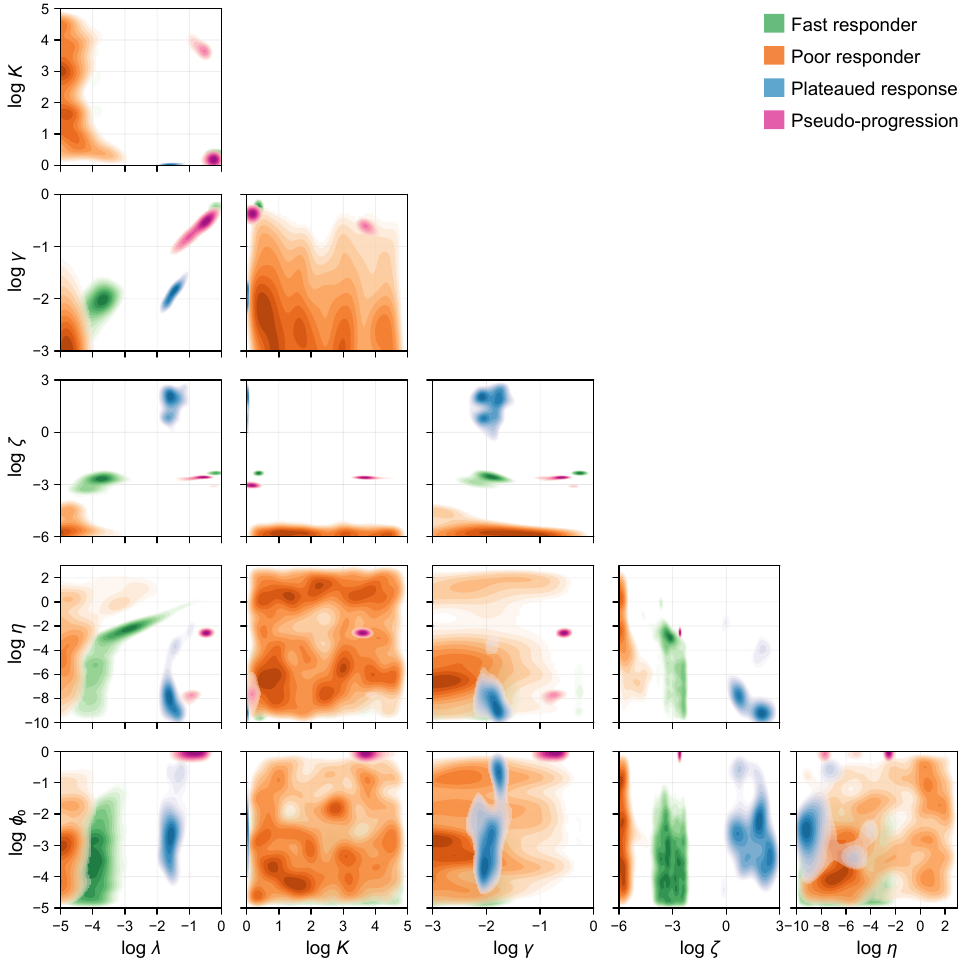}
		\caption[Figure 4]{\textbf{Parameter clustering according to classified patient response}. Kernel density of the full posterior distribution, following analysis of the training data set. Samples are classified into one of four patient responses according to criteria set out in \cref{sec:classification}, and kernel density estimates of bivariate marginal distributions conditioned on each classification shown. To aid comparison in the vicinity of the mode of each conditional posterior, only regions with densities greater than 50\% of the maximum are shown. Parameters relate to the cell proliferation rate, $\lambda$, the carrying capacity, $K$, the radiotherapy response strength, $\gamma$, the decay rate of necrotic debris, $\zeta$, the cell necrosis rate, $\eta$, and the initial proportion of the population that is necrotic, $\phi_0$. }
		\label{fig4}
	\end{figure}

The posterior for patient $i$ can be interpreted as the full posterior, conditioned on knowledge that the parameters relate to patient $i$
	\begin{equation}\label{individual_posterior}
		p^{(i)}(\bm\theta | \mathcal{D}_i) = p(\bm\theta | \{\mathcal{D}_i\}_{i=1}^n,i).
	\end{equation}
The \textit{full posterior} can be obtained by marginalising over all patients in the training set and is given by
	\begin{equation}\label{full_posterior}
		p(\bm\theta | \{\mathcal{D}_i\}_{i=1}^n) = \sum_{i} w_i \: p^{(i)}(\bm\theta | \mathcal{D}_i),
	\end{equation}
where $w_i = \mathbb{P}(i)$ represents the prior probability (i.e., weighting) of patient $i$. The result in \cref{full_posterior} follows immediately from \cref{individual_posterior} by the law of total probability. For simplicity, we set $w_i = \mathrm{const}$, however, such weights may be allowed to differ if additional knowledge informs patient similarity; for example, based on characteristics known to affect radiotherapy response, such as the clinical stage or age of a patient \cite{Belgioia.2021}. Another way to interpret the full posterior is that of a uniform mixture of the individual-level posterior distributions. We then denote the full posterior as the ``second-level prior'', $p_2(\bm\theta)$, which represents our knowledge about the parameters when analysing \textit{new} patients (we drop notational dependence on already observed data for convenience) (\cref{fig2}). An interpretation of our procedure is to identify the similarity between the new patient and the observed treatment outcomes for patients in the training set, and to combine the additional knowledge obtained from past patients when predicting outcomes for the new patient. In \cref{fig3} we compare the first-level prior $p_1(\bm\theta)$ to the full posterior, and in \cref{fig4} we show pairwise  marginal distributions of samples from the full posterior \cref{full_posterior}.

Given a possibly temporally incomplete set of measurements from a new patient, $\mathcal{D}_\text{new}$, the posterior distribution of the parameters is again given by
	\begin{equation}
		p_\text{new}(\bm\theta | \mathcal{D}_\text{new}) \propto p(\mathcal{D}_\text{new} | \bm\theta)\:p_2(\bm\theta).
	\end{equation}
A simple technique to obtain a set of weighted samples from $p_\text{new}(\bm\theta | \mathcal{D}_\text{new})$ is to apply a bootstrap particle filter to pre-obtained samples from $p_2(\bm\theta)$. Since patients in the training set are weighted equally, these may comprise a concatenation of samples from each posterior (we obtain these using an adaptive MCMC algorithm \cite{Vihola.2020uuo}, diagnostic statistics and convergence plots are given as supplementary material). An advantage of the bootstrap particle filter approach is that it requires minimal computational effort to update the posterior for new patients. The primary limitation introduced by this choice is that we cannot distinguish between parameters that vary between patients and those that are fixed: hence, we pre-estimate and fix the noise parameters in this work.

In practice, this approach may be problematic since patients in the training set are unlikely to be identically representative of new patients, particularly for small training sets (in our case, $n = 40$). In the bootstrap particle filter, this would lead to a small number of heavily weighted particles (that may or may not produce model realisations similar to the new patient data). We address this potential issue by forming $p_2(\bm\theta)$ by resampling perturbed particles from $p(\bm\theta | \{\mathcal{D}_i\}_{i=1}^n)$ using a multivariate normal distribution with covariance matrix, denoted $\Sigma_\varepsilon$, constructed by expanding the covariance matrix of Silverman's rule for kernel density estimation,
	\begin{equation}
		\Sigma_\varepsilon = \beta\left(\dfrac{4}{m(\dim(\bm\theta)+2)}\right)^{\frac{1}{\dim(\bm\theta)+4}} \mathrm{diag} \left(\Sigma_{\bm\theta}\right),
	\end{equation}
where $\beta$ is an expansion factor (we choose $\beta = 2$), $m$ is the number of samples of $\bm\theta | \{\mathcal{D}_i\}_{i=1}^n$ and $\Sigma_{\bm\theta}$ is the covariance matrix of the samples. We reject samples outside the support of the first-level prior $p_1(\bm\theta)$ (see \cref{tab2}), in effect constructing $p_2(\bm\theta)$ as a kernel density estimate with truncated multivariate normal kernels. This approach is also similar to a one-step sequential Monte Carlo algorithm \cite{Moral}.

\subsubsection{Quantifying goodness-of-fit}

We quantify goodness-of-fit using the so-called Bayesian $R^2$ statistic \cite{Gelman.2019}, defined for a single posterior sample by
	\begin{equation}
		R^2 = \dfrac{\text{Var}(V_\text{fit})}{\text{Var}(V_\text{fit}) + \text{Var}(V_\text{fit} - V_\text{obs})},
	\end{equation}
where $V_\text{fit}$ denotes the set of fitted values, and $V_\text{obs}$ denotes the set of observed values. A given posterior distribution yields a distribution of $R^2$ statistics: in this work, we report the median of the resultant distribution. Similarly to the frequentist $R^2$ statistic, a Bayesian $R^2$ statistic of unity indicates that the model captures all data variability (i.e., the variance of residuals, $\text{Var}(V_\text{fit} - V_\text{obs})$ is zero), while a Bayesian $R^2$ statistic of zero indicates that all fitted values lie on a horizontal line (hence, we expect low $R^2$ statistics for poor responders). 
	
\subsection{Generation of synthetic patient data}\label{synthetic_data_methods}

We generate synthetic patient data by resampling parameters from the full posterior and exposing patients to what we have previously referred to as a standard radiotherapy regime (weekday doses over a six week period, with CT measurements taken at the start of each treatment week and at the time of the final dose). Noise is added to synthetic measurements according to the statistical model (\cref{methods_stats}) with pre-estimated noise parameters. Synthetic data from a patient exhibiting a specific classification are produced by utilizing only full posterior samples that produce the classification of interest.

\section{Results and Discussion}

\subsection{Model calibration and patient classification}

To verify that the two compartment model can capture the range of radiotherapy responses observed \textit{in situ}, we first calibrate the untrained mathematical model to data from single patients in \cref{fig1}b--e using MCMC with the first-level prior. Best fits, along with the associated uncertainty in GTV, are shown alongside data in \cref{fig1}b--e. Overall, the model is able to reproduce clinical observations, although it has some difficulty distinguishing between fast responders and plateaued responses. Given that the plateaued response in \cref{fig1}d is diagnosed as such from only the last three observations, we attribute the potential for misclassification to uncertainty in the clinical observations (i.e., the noise model) and the lack of proceeding data points; it is impossible to tell whether this patient will continue to respond should treatment continue. Similar results are also seen for synthetic patients in the supplementary material, where patients that actually exhibit a plateaued response are classified as fast responders in the presence of noise (Fig. S2). 

Confident that the mathematical model can capture the observed range of responses, we proceed to train the model by sampling from the posterior for each of the 40 patients in the training set. The full posterior, formed by concatenating equal numbers of posterior samples from each patient in the training set (\cref{full_posterior}), is shown alongside the prior in \cref{fig3}. Note that the full posterior represents parameter combinations that can be attributed to patients throughout the population (the parameters vary patient-to-patient), and does not represent uncertainty in each parameter within any individual patient. Therefore, we are less interested in whether such parameters are identifiable, but rather that the full posterior now contains knowledge about the set of patient responses observed in the training set.

The correlation structure in the joint posterior is extremely important: marginal densities provide little information about each parameter and produce meaningless predictions when sampled independently. Therefore, in \cref{fig4} we investigate the correlation structure by examining the set of pair-wise bivariate marginal distributions. To gauge how parameter combinations vary with each radiotherapy response classification, we classify each posterior sample into a response class based on the criteria set out in \cref{sec:classification}. The proportion of samples attributed to each class is shown in \cref{tab1}. 

First, it is evident from results in \cref{fig4} that the predicted value of the initial necrotic proportion, $\phi_0$, does not vary between fast and poor responders. This is seen in bivariate denisties between $\phi_0$ and all other parameters. The statistic does, however, appear to distinguish pseudo-progressors from the other response types: estimates for $\phi_0$ suggest that tumours in such patients contain a much larger necrotic region pre-treatment. Faster responders are characterised in relation to poor responders by both a higher radiotherapy sensitivity, $\gamma$,  and necrotic material decay rate, $\zeta$. The necrotic material decay rate also appears to distinguish poor, fast, and plateaued responders: poor responders through a very low decay rate, plateaued responders by a high decay rate, and fast responders an intermediate rate. Finally, results in \cref{fig4} suggest that pseudo-progressors are characterised by both a high cell proliferation rate and correspondingly high radiotherapy response.

\subsection{Model predictions}

Given that the training set is relatively small, a potential obstacle is that responses of new patients may not be similar enough to those of existing patients to produce reliable predictions; indeed only 6.0\% of posterior samples correspond to patients that exhibit a poor response to treatment. To address this with the existing data, we ``expand'' the full posterior to form the second-level prior, $p_2(\bm\theta)$, by resampling and perturbing (essentially, forming $p_2(\bm\theta)$ as a multivariate kernel density estimate based on the full posterior, with a kernel variance expanded from Silverman's rule to account for new patient dissimilarity). The updated proportions, based on 100,000 samples from $p_2(\bm\theta)$, are given in \cref{tab1}, and suggest an updated prior probability of a new patient exhibiting a response at 65.3\%. An alternative approach that is beyond the scope of the current work would be to stratify perturbed full posterior samples based on an external and accepted classification ratio: for example, to choose the prior weights $\{w_i\}$ to achieve a desired prior ratio of patients in each classification. These results highlight the difficulty of classifying patient outcomes based on a relatively small cohort of patients with little prior parameter knowledge. Using the first-level prior (i.e., excluding all knowledge gained through analysis of the training data) further reduces the prior probability of an eventual response to 44.7\%.

	\begin{figure}[!b]
		\centering
		\includegraphics[width=\textwidth]{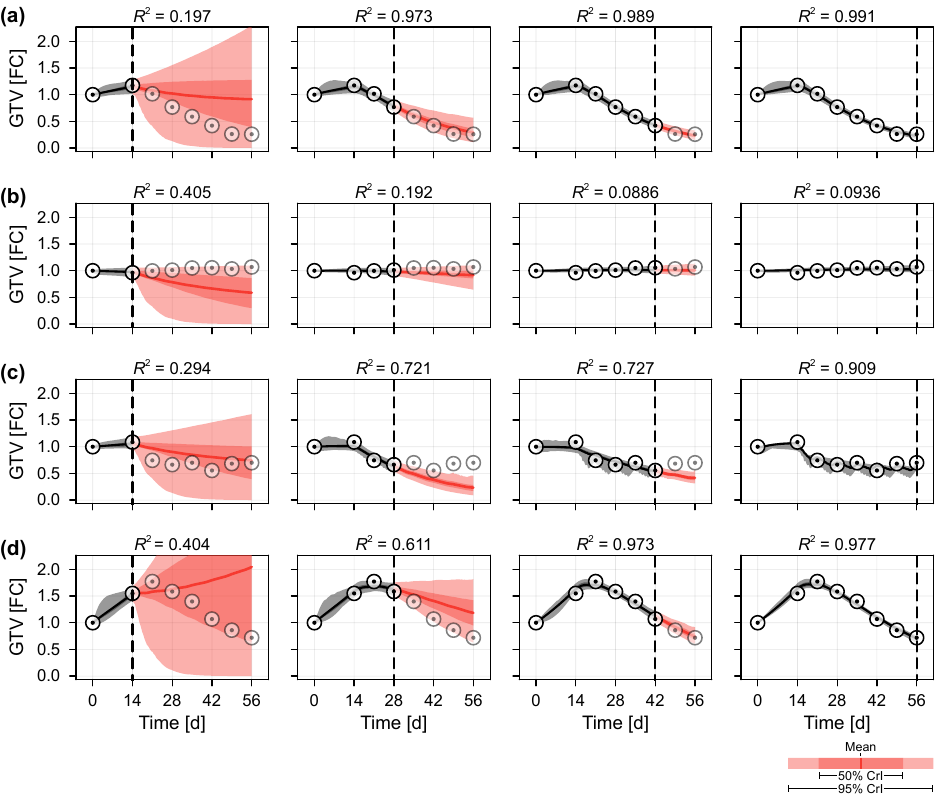}
		\caption[Figure 5]{\textbf{Temporal predictions for four synthetic patients.} Synthetic data from patients exhibiting (a) a fast response; (b) a poor response; (c) a plateaued response; and (d) pseudo-progression, are produced and used for predictions at various stages through the patient's treatment regime. In each case, the vertical dashed line indicates when the prediction is made: opaque marks indicate already-observed data used to produce predictions, semi-transparent marks indicate the future, as yet unobserved, trajectory. Predictions are represented as means (solid), 50\% credible intervals (dark black or red shading), and 95\% credible intervals (light black or red shading) constructed from weighted posterior samples. Model trajectories are coloured black (for retrospective predictions of tumour progression up to the present) and red (for prospective predictions of future tumour progression).}
		\label{fig5}
	\end{figure}

We first assess the predictive ability of our trained model by generating data from four synthetic patients exhibiting a fast response (\cref{fig5}a); a poor response (\cref{fig5}b); a plateaued response (\cref{fig5}c); and pseudo-progression (\cref{fig5}d). Given that each set of patient-specific parameters is resampled from the full posterior, we expect each synthetic patient to display a similar response to at least one patient in the training set. Additionally, as each set of synthetic data is generated by the mathematical model, we are guaranteed that the observed response is within the possible gamut of model responses.  We provide a table summarising the parameter values used for each patient in the supplementary material (Table S1).

	\begin{figure}
		\centering
		\includegraphics[width=\textwidth]{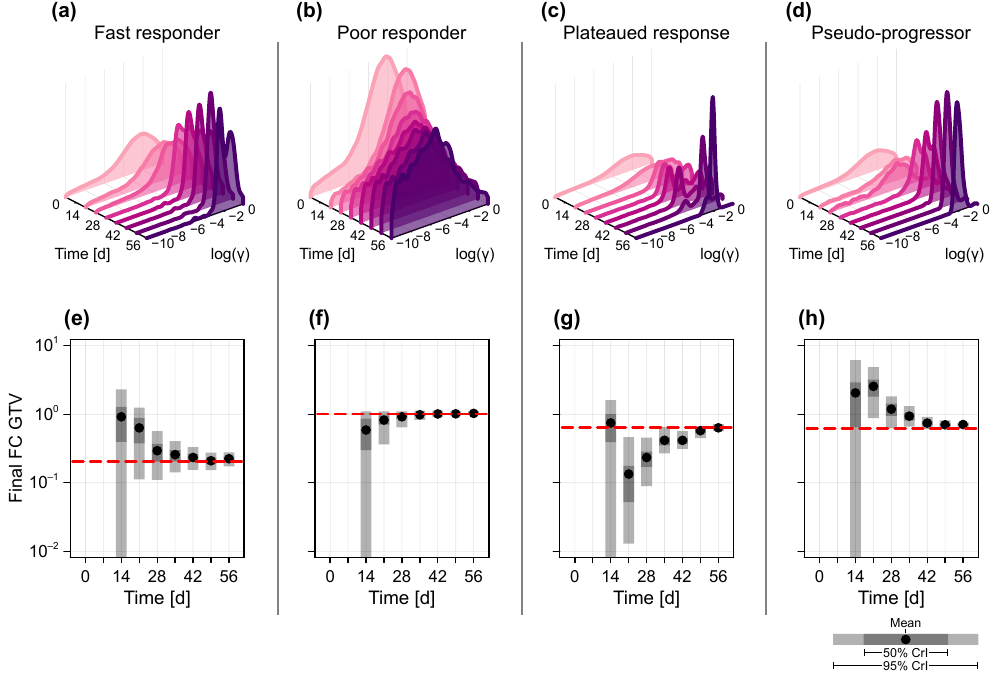}
		\caption[Figure 6]{\textbf{Predictions for four synthetic patients.} For the four patients analysed in \cref{fig5} we show (a--d) the evolution of the posterior distribution relating to radiotherapy response, $\gamma$; and (e--h) evolution of predictions for the relative tumour volume at the conclusion of treatment. In all cases, data up to, and including, the relevant time are included in the prediction. In (e--h), we show the mean (black disc), and both 50\% and 95\% credible intervals for the final tumour volume, together with the true final tumour volume (red dashed), both given as the fold-change (FC) relative to the initial volume, $V(0)$. The true values of $\gamma$ used for each patient are given as supplementary material (Table S1).}
		\label{fig6}
	\end{figure}

In \cref{fig5} we simulate real-time predictions by calibrating and forming predictions each week throughout treatment (i.e., at the time of each weekly CT scan). We show predictions made at the start of treatment ($t = \SI{14}{\day}$), and every second week following $(t = \;$\SIlist{28;42;56}{\day}). The results shown for $t = \SI{56}{\day}$ correspond to retrospective analysis of the trajectory, after all measurements have been taken, while the predictions drawn at $t = \SI{14}{\day}$ are made pre-treatment, before any radiotherapy response has been observed. As a class under-represented in the data set and hence the prior, predictions made for the pseudo-progressor at $t = \SI{28}{\day}$ almost entirely miss the true trajectory. Consequently, the single data point at $t = \SI{28}{\day}$ that sees a decrease is judged alongside both prior knowledge and potential measurement noise.

To quantitatively compare the time-evolution of prediction confidence, we plot in \cref{fig6}a--d the evolution of posterior information relating to the radiotherapy response, $\gamma$, and in \cref{fig6}e--h the time evolution of predicted final tumour volume (i.e., the fold-change GTV at $t = \SI{56}{\day}$ compared to the measurement at $t = \SI{0}{\day}$). The most immediate result is that both the fast responders and pseudo-progressors yield a posterior density for $\gamma$ higher than that for the poor-responders. The results in \cref{fig6}e show that the predicted final GTV quickly narrows around the true value for the fast responder, but takes longer for the plateaued progressors and pseudo-responders. At the same time, the results in \cref{fig6}f show that by two weeks into treatment, the model predicts with 95\% confidence that a patient will not see a final GTV less than 50\% of that pre-treatment. The results in \cref{fig6}g highlight again the difficulties faced when drawing predictions for patients exhibiting relatively rare responses: working with synthetic data eliminates the question of model-misspecification, however the 95\% credible intervals produced from predictions drawn at $t = \SI{21}{\day}$ and $t = \SI{28}{\day}$ do not cover the true value (which can be calculated by resimulating data from each synthetic patient without measurement noise). Given GTV alone, it is not until $t = \SI{42}{\day}$ (four weeks into treatment) that the model predicts with 95\% confidence that the patient's tumour will eventually see a reduction in volume. This is in line with previous reports that mid-treatment responses correlate with outcome \cite{Zahid.2021}.

\subsubsection{Clinical data}

Now that we have validated the model's ability to predict the time course of GTV for synthetic patients with a variety of radiotherapy responses, we turn to focus on drawing real-time predictions from unseen clinical data.

	\begin{figure}[!b]
		\centering
		\includegraphics[width=\textwidth]{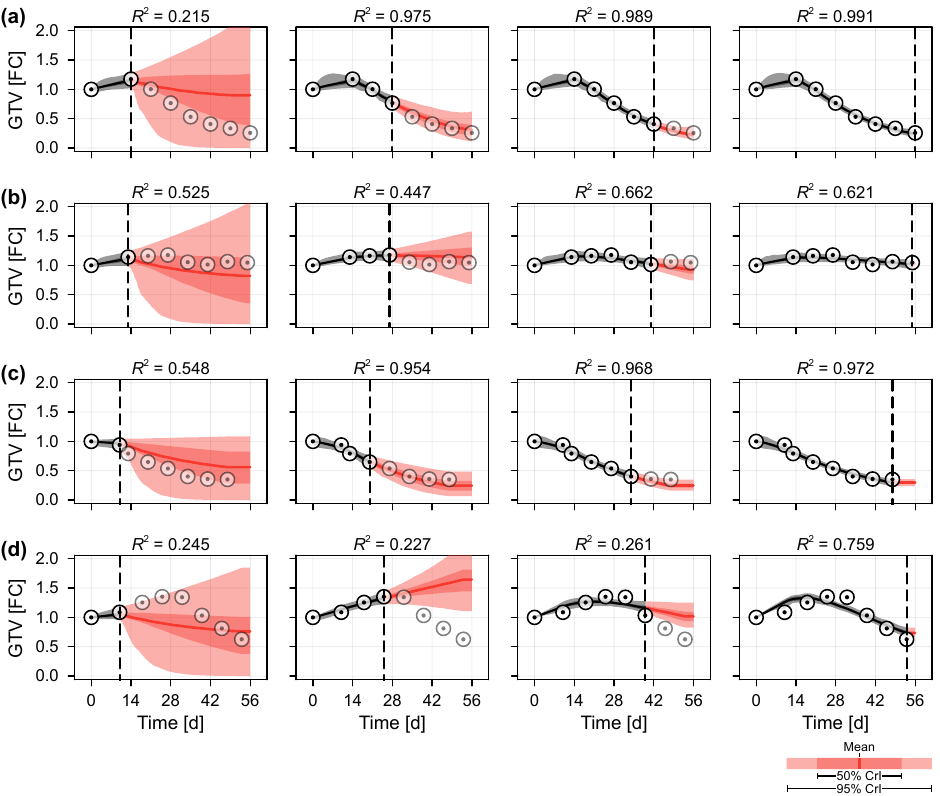}
		\caption[Figure 7]{\textbf{Temporal predictions for the four patients excluded from the training set.} We reproduce the analysis from \cref{fig5} for the four patients in \cref{fig1}. These patients were not included in the training set, and so these results are representative of clinical predictions made throughout a new patient's course of treatment. Patients were classified previously as (a) a fast responder; (b) a poor responder; (c) exhibiting a plateaued response; and, (d) exhibiting pseudo-progression. Results related to the remaining seven patients excluded from the training set are given in the supplementary material (Fig. S6).}
		\label{fig7}
	\end{figure}
	
	\begin{figure}[!t]
		\centering
		\vfill
		\includegraphics[width=\textwidth]{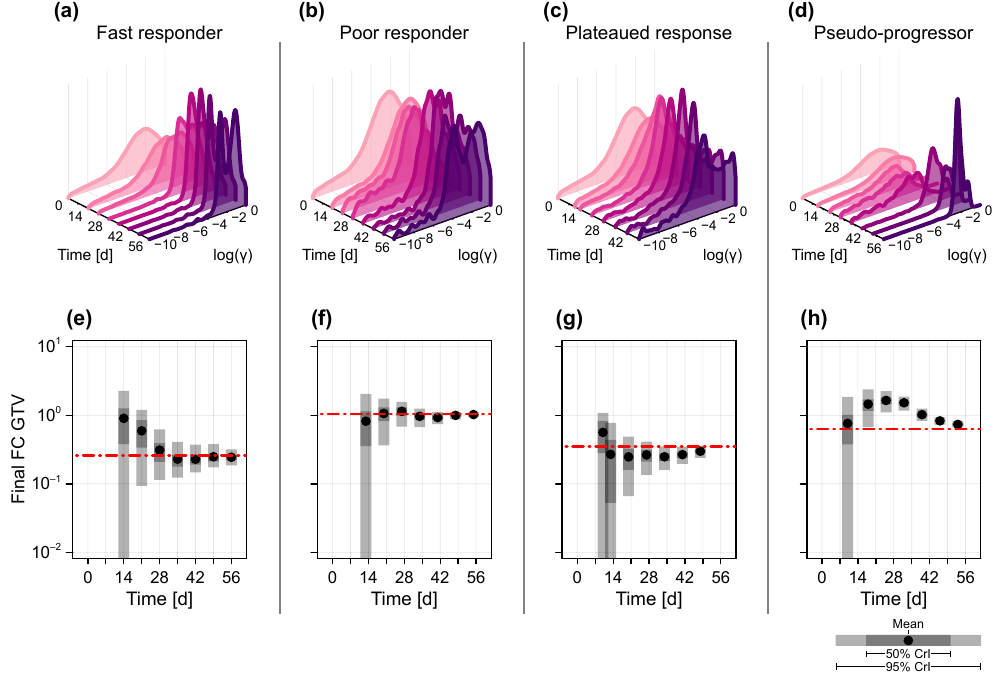}
		\caption[Figure 8]{\textbf{Predictions for the four patients excluded from the training set.} We reproduce the analysis from \cref{fig6} for the four patients in \cref{fig1}b--e. These patients were not included in the training set, and so these results are representative of clinical predictions made throughout a new patient's course of treatment.}
		\label{fig8}
		\vfill
	\end{figure}

	In \cref{fig7,fig8}, we repeat the analysis performed in \cref{fig5,fig6} for the four patients initially exhibited in \cref{fig1}. We remind the reader that, although we previously demonstrated that the model can reproduce the clinical observations for these four patients, none were included in the training set. Hence, predictions drawn up to a particular time include only GTV data up to and including that time, and knowledge gained from the training set. For completeness, in the supplementary material we reproduce the results in \cref{fig7} for all 51 patients using a leave-out-one-cross-validation approach, where predictions for each patient are drawn from a training set comprising the other 50 patients.
	
	At the time of treatment onset ($t = \SI{14}{\day}$ in \cref{fig7}a,b and $t = \SI{12}{\day}$ in \cref{fig7}c,d), predicted trajectories are similar and predominantly represent prior knowledge from the training set. By day 28, for the fast responder, and day 21, for the patient that eventually exhibits a plateaued response, the model predicts with 95\% confidence that the patient will eventually achieve an overall reduction in tumour volume. Indeed, for both of these patients the precision in predictions of the final tumour volume narrows quickly around what is eventually observed. In contrast, at day 28 the patient that eventually exhibits a poor response sees roughly half of all predicted trajectories indicating an eventual increase in volume, and half a decrease. Throughout treatment, the mean prediction remains around the eventually observed value of unity. The results for the pseudo-progressor mirror those observed in the synthetic data: the predictions are perhaps initially misleading due to the relatively small (2.3\%) prior probability of a patient exhibiting such a response.

\begin{figure}
		\centering
		\includegraphics{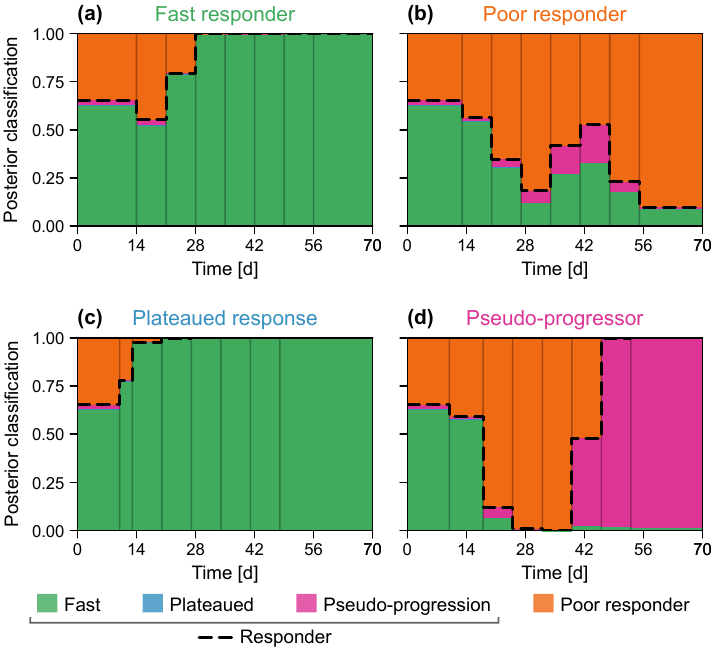}
		\caption[Figure 9]{\textbf{Classification of the four patients excluded from the training set.} We predict each patient's classified response using data up to and including the relevant time (height of each region indicates the predicted proportion). The predicted probability of the patient responding (i.e., receiving a classification that is not that of a poor responder) is shown in black dashed. Before the start of treatment, the predicted classifications correspond to those of the second-level prior in \cref{tab2}. }
		\label{fig9}
	\end{figure}

	To quantitatively explore the model's ability to predict patient classification, in \cref{fig9} we plot the posterior classification probabilities for predictions drawn at each time point, in addition to a pooled classification probability of a patient displaying a response (i.e., not a poor responder). Initially, at $t = \SI{0}{\day}$, the classification probabilities represent those in the second-level prior, $p_2(\bm\theta)$ (\cref{tab1}). The most notable results are for the relatively rare classifications of plateaued response and pseudo-progressor. In the case of the former, the patient has a posterior classification mode (i.e., the most likely classification given all the information collected during the patients' course of treatment) of a fast responder. This again highlights the difficulties distinguishing plateaued responses from observation noise seen in faster responders. The pseudo-progressor, however, begins to gain a correct posterior classification probability by $t = \SI{42}{\day}$, just over four weeks into treatment. The classifications following the first measurement at $t = \SI{14}{\day}$ are qualitatively similar to that observed in the prior, subsequent measurements which show an increase in gross tumour volume lead to classification as a poor responder, highlighting the limitations of the currently trained model in distinguishing pseudo-progressors from poor responders.

	To explore the relative value of existing and newly collected information, in the supplementary material we produce additional results that show temporal predictions for both synthetic and validation patients, produced using the uninformative (i.e., first-level) prior. These results correspond to a prior probability of 44.7\% that a patient will eventually respond to treatment; much lower than that estimated from analysis of the training data (94.0\%) and that in the second-level prior (65.3\%). For the synthetic patients presented in \cref{fig5}, the results show a decrease in prediction fit (as measured by Bayesian $R^2$) for predictions drawn prior to $t = \SI{28}{\day}$. For times later than $t = \SI{42}{\day}$, predictions drawn using both the uninformative and informative priors are comparable. Similar results are seen for the validation patients presented in \cref{fig7}, although the differences are less pronounced from the third-post-radiotherapy observation point onwards. The difference between results for the synthetic and validation patients is expected: the informative prior is known to be representative for the synthetic patients, whereas we do not have this guarantee for the validation patients. Hence, observed information is more important than prior information in newly informed patients that are not well-represented by the prior.

\subsection{Value in collecting measurements of tumour heterogeneity}

\begin{figure}[h]
		\centering
		\includegraphics[width=\textwidth]{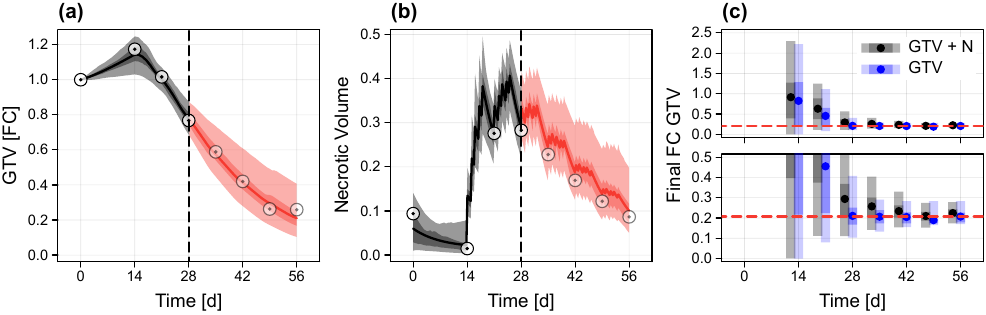}
		\caption[Figure 10]{\textbf{Predictions for a synthetic patient with a fast response subject to both GTV and necrosis measurements.} (a--b) We reproduce the analysis from \cref{fig5}a in the case that information relating to both $V(t)$ and $N(t)$ is available. (c) Mean, 50\%, and 95\% credible intervals for the final GTV in both data collection scenarios. The true value (calculated by resimulating data from each synthetic patient without measurement noise) is also shown (red dashed). Lower plot in (c) is a cropped inset of the upper.}
		\label{fig10}
	\end{figure}

The weekly GTV used for our analysis already exceeds clinical practice of just two pretreatment CT scans per patient. To assess the potential value of collecting higher-quality scan data that additionally enables identification of the tumour's necrotic volume, we repeat our analysis of the synthetic patient in \cref{fig5}a given that noisy measurements of both $V(t)$ and $N(t)$ are now available. The results in \cref{fig10}a,b show that, by day 28, relatively precise predictions relating to the trajectories of both variables can now be made. In \cref{fig10}c we quantitatively compare predictions for the final GTV in both scenarios. As expected, more precise estimates can be made should data relating to both variables be available.

\begin{figure}[h]
		\centering
		\includegraphics[width=\textwidth]{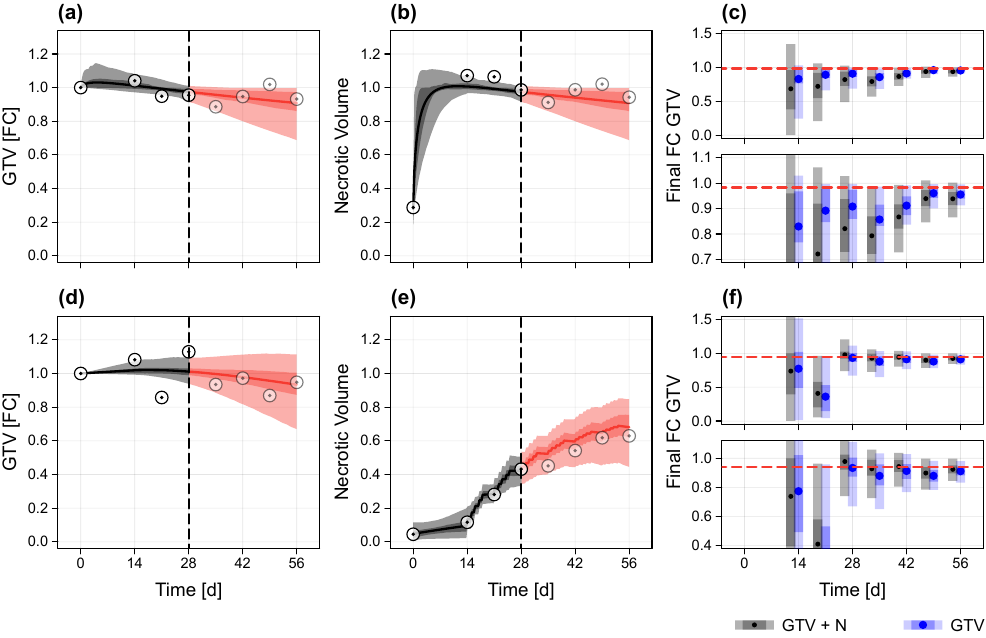}
		\caption[Figure 11]{\textbf{Predictions for two synthetic patients with a poor response subject to both GTV and necrosis measurements.} (a--b,d--e) We produce dynamic predictions of tumour progression for each patient in the case that information relating to both $V(t)$ and $N(t)$ is available. (c,f) Mean, 50\%, and 95\% credible intervals for the final GTV in both data collection scenarios. The true value (calculated by resimulating data from each synthetic patient without measurement noise) is also shown (red dashed). Lower plot in each set is a cropped inset of the upper.}
		\label{fig11}
	\end{figure}

In \cref{fig11}, we repeat the analysis for two new synthetic patients that experience a poor response. In the case of the first patient, a small gain in predictive ability is seen from the inclusion of necrotic volume measurements (\cref{fig11}c); interestingly, this improvement is not seen for the second patient (\cref{fig11}f). Overall, these results highlight a key challenge with using the population-calibrated mathematical model to draw predictions relating to tumour composition and the underlying cause of a poor response, particularly given the wide-ranging spatial compositions seen in poor responders. The first synthetic patient exhibits a poor response due to the development of a tumour comprising almost entirely necrotic material, which does not degrade (\cref{fig11}b), while the tumour composition in the second synthetic patient is perhaps more realistic, with the necrotic fraction comprising approximately 60\% of the GTV at the end of treatment. (\cref{fig11}e). Since the model is not trained using clinical data relating to tumour composition, it cannot distinguish between tumour compositions that are clinically realistic and those that are not. This is not an issue for prediction of the GTV, as prediction uncertainty incorporates all possible tumour compositions through prior knowledge. Predictions of necrotic volume, meanwhile, represent predominantly prior knowledge in addition to restrictions imposed by the modelled relationship between the observed GTV of patients in the training set and their potential inner tumour composition.

\section{Conclusion}

The development of predictive mathematical models of patient-specific tumour response is hindered by multiple challenges. Mathematical models must incorporate sufficient detail to capture a wide range of potential responses, while clinical data are highly limited, often comprising just one or two noisy measurements of tumour volume prior to treatment initiation. Advances in imaging technologies or the use of magnetic resonance imaging embedded in radiation delivery devices may, in future, provide a cost-effective means of collecting more detailed information, allowing the calibration of correspondingly more detailed mathematical models \cite{Gatenby.2013,Gillies.2018,McGee.2021,Park.2023}. In this work, however, we work with a fundamental set of measurements, and present the statistical methodology and an appropriately complex mathematical model to maximise data utility and draw clinically relevant predictions by leveraging a cohort of patients that exhibit a variety of treatment responses.

Importantly, the two compartment model is able to reproduce the full range of patient responses observed in our cohort of clinical data, representing an improvement over previously proposed one-compartment models which may not capture more complex behaviours, such as the plateaued response and pseudo-progressor behaviour. This is particularly important for prediction, since the choice of model and gamut of possible responses form a significant part of prior knowledge. While the mathematical literature presents an extensive catalogue of more complex models, we find that our choice of model with six unknown parameters, all with a direct biophysical interpretation, is simultaneously both sufficiently simple to ensure practical identifiability in some cases, and sufficiently complex to produce the variety of responses seen in the clinical data. Parameter identifiability is clearly not essential to produce predictions (single patient predictions drawn early in the course of treatment from the first-level prior, where the number of parameters exceeds the number of data points, are still sensical), however the relatively small parameter space and resultant tightly constrained second-level prior (\cref{fig4}) ensures adequate coverage in our resampling-based inference method: we expect our approach to  become prohibitively expensive for models with large numbers of parameters.

The overarching goal of the presented framework is to leverage existing clinical data to produce a predictive model for GTV that accurately captures the uncertainty in predictions made for new patients. By benchmarking against both synthetic and a validation clinical data set, we show that our approach excels at this goal for patients with more typical responses: the fast and poor responders. Given the relatively small size of our training data---comprising measurements from 40 patients---it is no surprise that our approach does not perform as well for patients with atypical responses: pseudo-progressors, for instance, make up only 2.3\% of the prior, meaning that the GTV progression of these patients is informed by (on average) a single patient in the training set. In this case, it takes six on-treatment measurements before the patient is identified as more likely to exhibit an eventual response than a poor response. The most effective remedy would be to accumulate significantly more clinical data with better representation of outliers. Should enough data become available, stratification could be used to ensure that representation of patients in the training data either concords with that in the population, or incorporates non-quantitative prior knowledge (such as patient characteristics) that pre-inform similarities with patients in the training set. Our modelling framework is well-poised to incorporate more detailed clinical data, including, for instance, radiotherapy plan adaptation and information relating to variations in delivered dose throughout the course of treatment. Inclusion of such information is likely to lead to better response classification, particularly if the radiotherapy dose is modified during the course of treatment.

Both the accuracy and precision of predictions could also be improved for all patients through a better biological understanding of radiotherapy response. The final set of results presented in this work highlight that GTV measurements alone are insufficient to identify the root cause of a poor response. Indeed, predictions related to the inner tumour composition must be treated with as much caution as with predictions for atypical patients that are dissimilar to all patients in the training set. The absence of tumour composition data in the training set means that all predictions of tumour composition are only informed by data indirectly through the model, which has, in turn, been validated against solely GTV data. The prospect of training a model with joint GTV-composition measurements is at present hypothetical, although entirely possible through advanced imaging technologies \cite{Sun.2018,Salem.2019,Rockne.2019}. At this stage, our framework could additionally be applied to answer important questions relating to the number of tumour composition measurements required to accurately predict patient outcome throughout their course of treatment.

We highlight that, in general, our statistical methodology is entirely model agnostic. Thus, informed by more detailed data, our approach could be used to develop a fully validated predictive model of not just GTV, but tumour composition, cell density, proliferation, hypoxia, and more. However, this proposition is not without limitation: our current choice to bootstrap parameter samples is likely to perform poorly for models with a large number of parameters. Such dimensionality-induced issues can be in part alleviated by sampling the full posterior directly, although this would introduce additional computational challenges. Further statistical developments are also needed to include parameters that are fixed between patients (for example, the noise parameters), or parameters that are assumed to be uncorrelated to others.

Our results add to a growing body of work \cite{Claret.2009,Ribba.2012,Rockne.2019,Bruno.2020} that highlights the utility that mathematical models could bring to the clinic; in future informed by highly detailed and representative patient data to provide objective, real-time, and personalised patient predictions that inform clinical decision-making.

\section*{Data availability}
	Code used to produce the results are available on GitHub at \url{https://github.com/ap-browning/clinical_predictions}.

\section*{Author contributions}
	All authors conceived the study, provided feedback on drafts, and gave approval for final publication. A.P.B. and T.L. drafted the manuscript. A.P.B. implemented the computational algorithms. J.C. collected and provided the clinical data.

\section*{Acknowledgements}
	This work was funded in part by the Engineering and Physical Sciences Research Council (grant number EP/G037280/1). T.L. would also like to thank the Moffitt Cancer Center, where some of this work was undertaken, for their hospitality.
	


\end{document}


	\title{Supplementary material for \\ ``Predicting radiotherapy patient outcomes with real-time clinical data using mathematical modelling''}


	\author[1]{Alexander P Browning\textsuperscript{*}\textsuperscript{$\ddagger$}}
	\author[1,2]{Thomas D Lewin\textsuperscript{*}}
	\author[1]{Ruth E Baker}
	\author[1]{Philip K Maini}
	\author[3]{Eduardo G. Moros}
	\author[3]{Jimmy Caudell}
	\author[1]{Helen M Byrne\textsuperscript{$\dagger$}}
	\author[3,4,5]{Heiko Enderling\textsuperscript{$\dagger$}}

	\affil[1]{Mathematical Institute, University of Oxford, Oxford, UK}
	\affil[2]{Roche Pharma Research and Early Development, Roche Innovation Center, Basel, Switzerland}
	\affil[3]{Department of Radiation Oncology, H. Lee Moffitt Cancer Center \& Research Institute, USA}
	\affil[4]{Department of Integrated Mathematical Oncology, H. Lee Moffitt Cancer Center \& Research Institute, USA} 
	\affil[5]{New Address: Department of Radiation Oncology, MD Anderson Cancer Center, Houston, TX, USA}


\date{\today}
\maketitle

\footnotetext[1]{These authors contributed equally.}
\footnotetext[2]{These authors also contributed equally.}
\footnotetext[3]{Corresponding author: browning@maths.ox.ac.uk or HEnderling@mdanderson.org}

\renewcommand\thefigure{S\arabic{figure}} 
\renewcommand\thetable{S\arabic{table}} 

	\begin{table}[H]
		\centering
		\caption{Parameters resampled from the full posterior corresponding to four synthetic patients.}
		\label{tabS1}
		\begin{tabular}{ccccc}\hline
			Parameter 	& Fast responder 		& Poor responder		& Plateaued responder 	& Pseudo progression \\\hline\hline
			$\lambda$ 	& \SI{3.5e-2}{\per\day} & \SI{3.0e-1}{\per\day} & \SI{3.2e-1}{\per\day}	& \SI{3.2e-1}{\per\day}\\
			$K$ 			& \SI{2.2}{} 			& \SI{2.2}{} 			& \SI{1.1}{} 			& \SI{1.5}{}\\
			$\gamma$ 	& \SI{1.1e-1}{\per\day} & \SI{1.0e-2}{\per\day} 	& \SI{2.2e-1}{\per\day}	& \SI{4.2e-1}{\per\day}\\
			$\zeta$ 		& \SI{1.4e-1}{\per\day} & \SI{2.4e-4}{\per\day} & \SI{3.6}{\per\day} 	& \SI{4.5e-2}{\per\day}\\
			$\eta$ 		& \SI{1.6e-4}{\per\day} & \SI{8.3}{\per\day} 	& \SI{1.3e-4}{\per\day} & \SI{1.4e-4}{\per\day}\\
			$\phi_i$ 	& \SI{9.4e-2}{} 		& \SI{1.5e-1}{} 		& \SI{1.5e-1}{}			& \SI{2.9e-1}{}\\\hline
		\end{tabular}
	\end{table}

	\begin{figure}
		\centering
		\includegraphics[width=\textwidth]{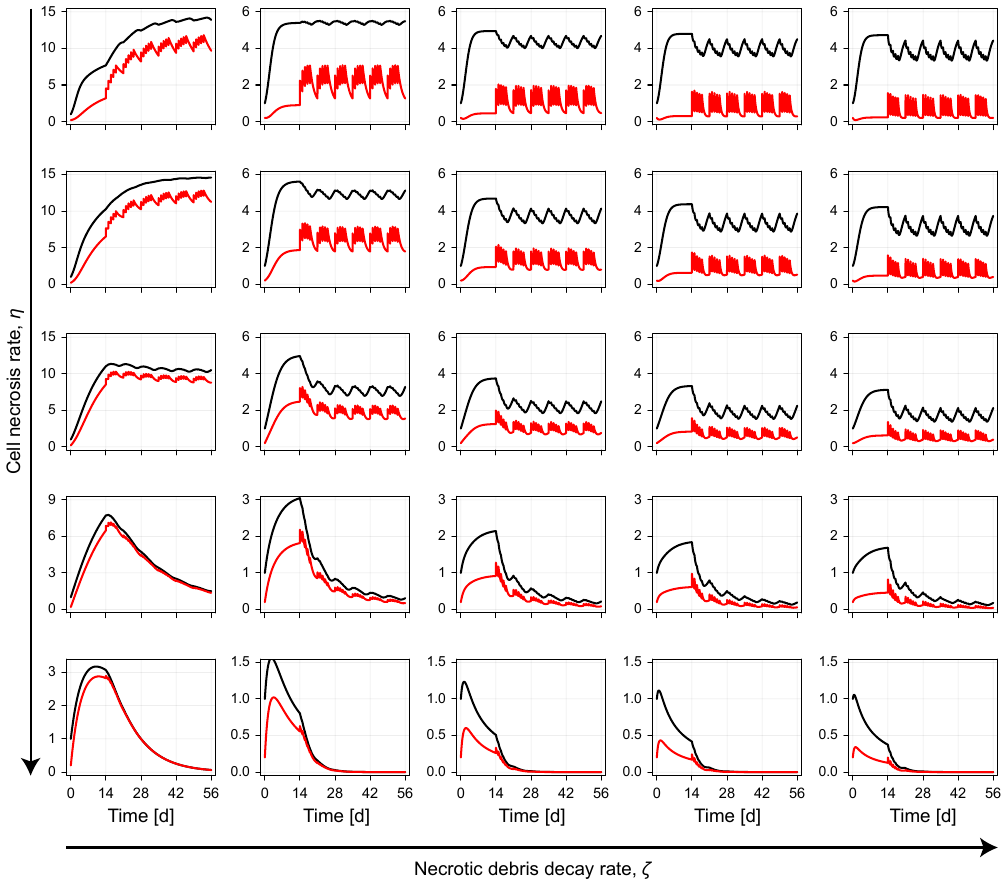}
		\caption[Figure S1]{\textbf{Two parameter sweep of mathematical model.} We sweep across the parameters $\zeta \in \{0.1,0.5,1,1.5,2\}$ and $\eta \in \{0.1,0.25,0.5,0.75,1\}$ with the other parameters fixed at $\lambda = 1$, $K = 5$, $\gamma = 0.3$, and $\phi_0 = 0.2$. Arrows indicate the direction of increasing $\zeta$ and $\eta$. Trajectories show the total tumour volume (black), and necrotic volume (red). All patients undergo the standard course of treatment used in the classification procedure in the main document (daily doses of radiotherapy on weekdays over a six week period, initiated from $t = \SI{14}{\day}$).}
		\label{figS1}
	\end{figure}
	
	\begin{figure}[h]
		\centering
		\includegraphics{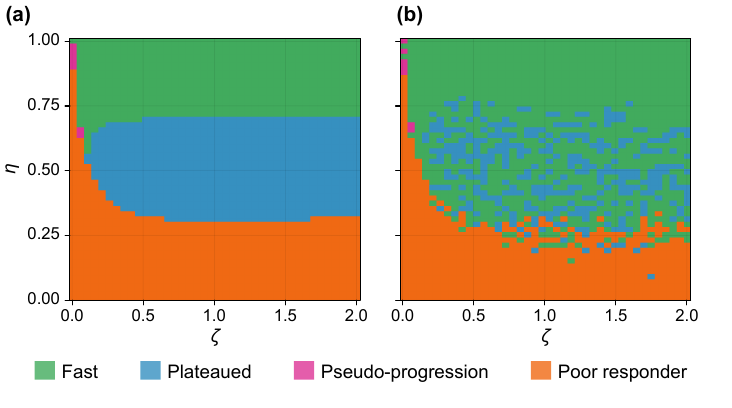}
		\caption[Figure S2]{\textbf{Classified patient responses from two parameter sweep.} We sweep across the parameters $0 < \zeta \le 2$ and $0 < \eta \le 1$ with the other parameters held fixed at $\lambda = 1$, $K = 5$, $\gamma = 0.3$, and $\phi_0 = 0.2$. All patients undergo the standard course of treatment used in the classification procedure in the main document (daily doses of radiotherapy on weekdays over a six week period, initiated from $t = \SI{14}{\day}$). In (a), classification is performed using noise-free synthetic data, whilst in (b), classification is performed using noisy synthetic data produced using the statistical model with pre-estimated noise parameters.}
		\label{figS2}
	\end{figure}

	\begin{table}[h]
\centering
\label{tabS2}
\caption{$\hat{R}$ MCMC diagnostic statistics for each patient.}
\begin{tabular}{cccccccc}\hline
ID & In Training & $\lambda$     & $K$     & $\gamma$     & $\zeta$    & $\eta$    & $\phi_0$    \\\hline\hline
1  & Yes         & 1.003 & 1.001 & 1.004 & 1.005 & 1.002 & 1.003 \\
2  & Yes         & 1.001 & 1.002 & 1.0   & 1.0   & 1.001 & 1.002 \\
3  & Yes         & 1.025 & 1.017 & 1.003 & 1.007 & 1.007 & 1.007 \\
4  & Yes         & 1.0   & 1.001 & 1.003 & 1.001 & 1.002 & 1.002 \\
5  & No        & 1.067 & 1.051 & 1.063 & 1.021 & 1.029 & 1.047 \\
6  & No        & 1.0   & 1.003 & 1.0   & 1.0   & 1.0   & 1.001 \\
7  & Yes         & 1.001 & 1.003 & 1.001 & 1.001 & 1.001 & 1.001 \\
8  & Yes         & 1.003 & 1.004 & 1.001 & 1.001 & 1.001 & 1.0   \\
9  & Yes         & 1.002 & 1.001 & 1.002 & 1.002 & 1.001 & 1.003 \\
10 & Yes         & 1.006 & 1.004 & 1.003 & 1.001 & 1.008 & 1.008 \\
11 & Yes         & 1.008 & 1.002 & 1.009 & 1.003 & 1.008 & 1.006 \\
12 & Yes         & 1.001 & 1.0   & 1.007 & 1.002 & 1.002 & 1.001 \\
13 & No        & 1.011 & 1.017 & 1.015 & 1.002 & 1.002 & 1.014 \\
14 & No        & 1.005 & 1.004 & 1.004 & 1.001 & 1.005 & 1.002 \\
15 & Yes         & 1.006 & 1.005 & 1.007 & 1.003 & 1.009 & 1.002 \\
16 & Yes         & 1.001 & 1.002 & 1.001 & 1.002 & 1.002 & 1.003 \\
17 & Yes         & 1.001 & 1.001 & 1.001 & 1.001 & 1.001 & 1.003 \\
18 & Yes         & 1.0   & 1.0   & 1.0   & 1.0   & 1.002 & 1.0   \\
19 & Yes         & 1.004 & 1.003 & 1.009 & 1.006 & 1.004 & 1.001 \\
20 & Yes         & 1.011 & 1.009 & 1.009 & 1.011 & 1.008 & 1.025 \\
21 & No        & 1.001 & 1.002 & 1.0   & 1.003 & 1.001 & 1.001 \\
22 & Yes         & 1.008 & 1.003 & 1.007 & 1.004 & 1.012 & 1.004 \\
23 & Yes         & 1.006 & 1.0   & 1.007 & 1.003 & 1.009 & 1.005 \\
24 & Yes         & 1.0   & 1.001 & 1.0   & 0.999 & 1.001 & 1.001 \\
25 & No        & 1.001 & 1.001 & 1.0   & 1.001 & 1.003 & 1.002 \\
26 & Yes         & 1.002 & 1.001 & 1.001 & 1.003 & 1.002 & 1.001 \\
27 & Yes         & 1.001 & 1.0   & 1.0   & 1.0   & 1.0   & 1.0   \\
28 & Yes         & 1.003 & 1.007 & 1.006 & 1.015 & 1.004 & 1.008 \\
29 & No        & 1.003 & 1.001 & 1.0   & 1.001 & 1.002 & 1.001 \\
30 & Yes         & 1.002 & 1.002 & 1.003 & 1.001 & 0.999 & 1.0   \\
31 & Yes         & 1.0   & 1.0   & 1.001 & 1.001 & 1.001 & 1.0   \\
32 & Yes         & 1.001 & 1.001 & 1.0   & 1.0   & 1.0   & 1.001 \\
33 & Yes         & 1.003 & 1.001 & 1.0   & 1.001 & 0.999 & 1.002 \\
34 & No        & 1.001 & 1.002 & 1.002 & 1.002 & 1.0   & 1.001 \\
35 & Yes         & 1.01  & 1.001 & 1.003 & 1.007 & 1.009 & 1.008 \\
36 & No        & 0.999 & 1.001 & 1.0   & 1.001 & 1.001 & 1.001 \\
37 & Yes         & 1.004 & 1.004 & 1.001 & 1.002 & 1.003 & 1.004 \\
38 & Yes         & 1.0   & 1.001 & 1.006 & 1.001 & 1.003 & 1.001 \\
39 & Yes         & 1.004 & 1.004 & 1.003 & 1.0   & 1.003 & 1.002 \\
40 & Yes         & 1.013 & 1.004 & 1.01  & 1.001 & 1.004 & 1.008 \\
41 & No        & 1.004 & 1.005 & 1.009 & 1.009 & 1.003 & 1.003 \\
42 & Yes         & 1.002 & 1.0   & 1.001 & 1.001 & 1.001 & 1.001 \\
43 & Yes         & 1.003 & 1.003 & 1.001 & 1.002 & 1.0   & 1.003 \\
44 & Yes         & 1.0   & 1.001 & 1.0   & 1.001 & 1.001 & 1.0   \\
45 & Yes         & 1.134 & 1.208 & 1.1   & 1.121 & 1.129 & 1.296 \\
46 & Yes         & 1.001 & 1.0   & 1.0   & 1.002 & 1.003 & 1.001 \\
47 & Yes         & 1.002 & 1.004 & 1.008 & 1.001 & 1.009 & 1.001 \\
48 & Yes         & 1.43  & 1.575 & 1.447 & 1.781 & 1.715 & 1.537 \\
49 & Yes         & 1.001 & 1.0   & 1.0   & 1.001 & 1.0   & 0.999 \\
50 & Yes         & 1.003 & 1.002 & 1.008 & 1.001 & 1.008 & 1.004 \\
51 & No        & 1.002 & 1.0   & 1.001 & 1.003 & 1.0   & 1.0 \\\hline
\end{tabular}
\end{table}

	\begin{figure}[h]
		\centering
		\includegraphics[width=0.9\textwidth]{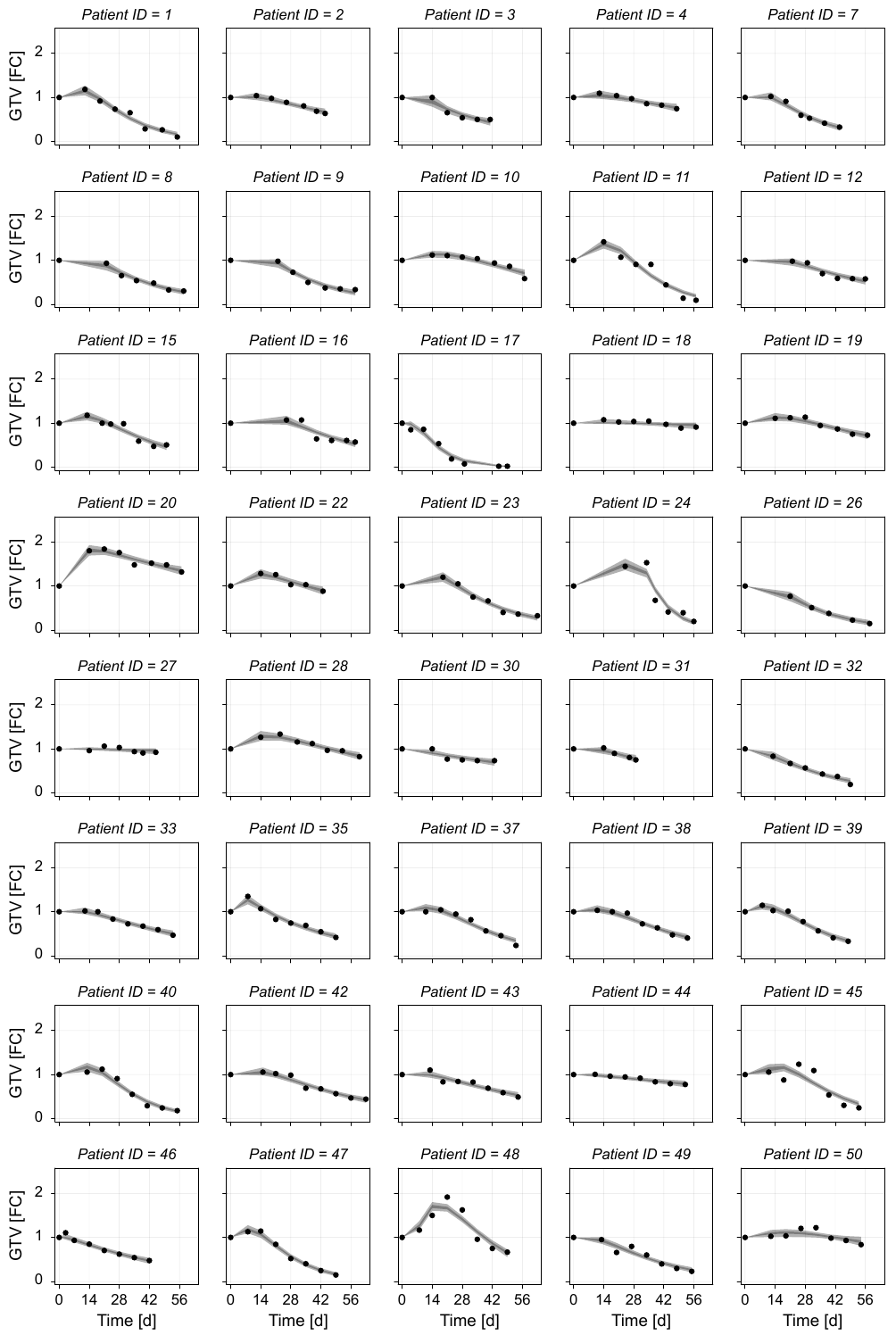}
		\caption[Figure S3]{ \textbf{Fits for patients in the training set.} Individual fits for all patients in the training set, using the population-level prior. Shown are the data (black disc), 50\% credible interval (light grey), and 95\% credible interval (dark grey). Note that model predictions are only drawn at times corresponding to clinical measurements: results for intermediate time points show as a linear interpolation.}
		\label{figS3}
	\end{figure}

	\begin{figure}[h]
		\centering
		\includegraphics{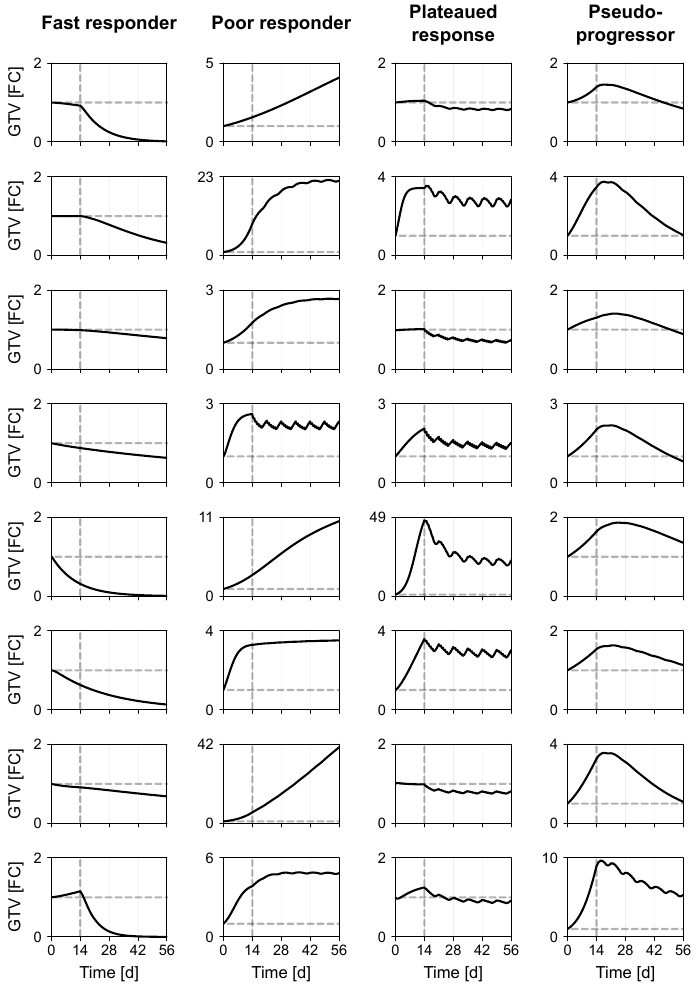}
		\caption[Figure S4]{\textbf{Subset of classified prior samples.} We demonstrate the choices in the classification algorithm by simulating eight patients of each class from the prior distribution. All patients undergo the standard course of treatment. Shown is the tumour volume (black), a horizontal line indicating unity (dashed grey), and a vertical line showing the time of the first radiotherapy dose (dashed grey). }
		\label{figS4}
	\end{figure}

	\begin{figure}[h]
		\centering
		\includegraphics[width=\textwidth]{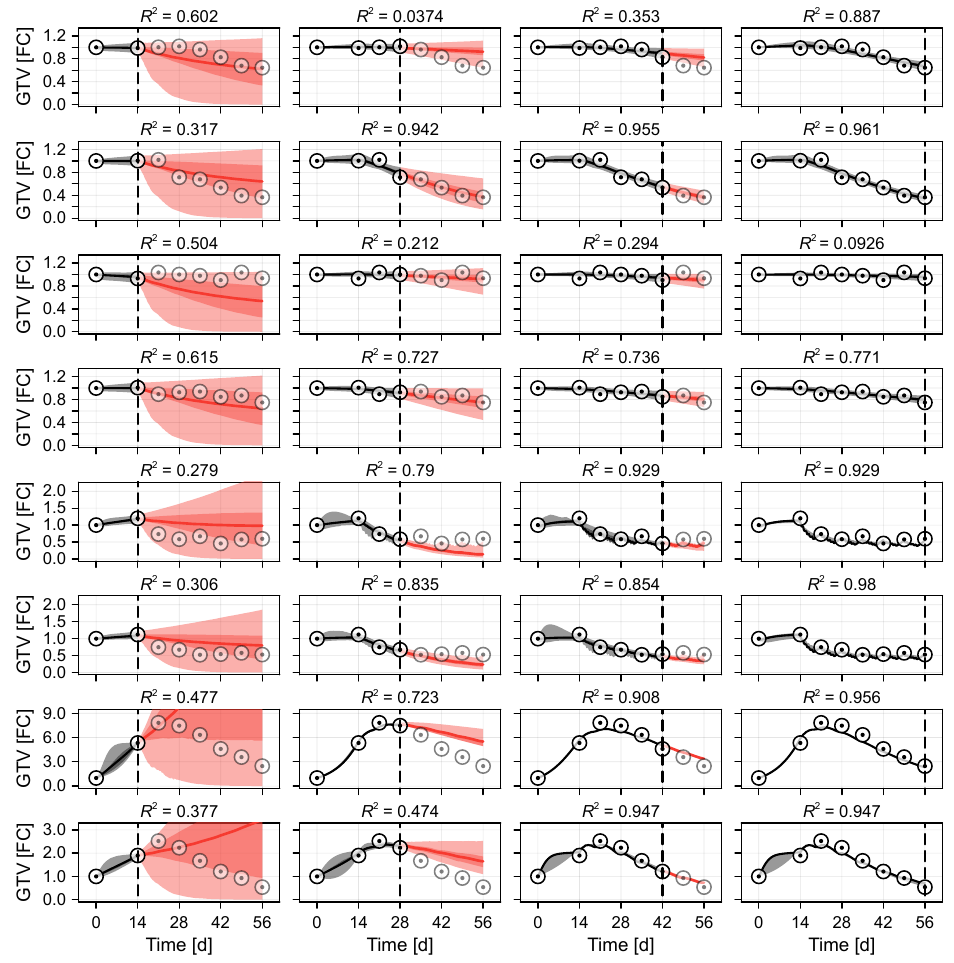}
		\caption[Figure S5]{ \textbf{Results for eight additional synthetic patients.} We reproduce the analysis from Fig. 5 in the main document for eight additional synthetic patients (two from each response classification). }
		\label{figS5}
	\end{figure}

	\begin{figure}[h]
		\centering
		\includegraphics[width=\textwidth]{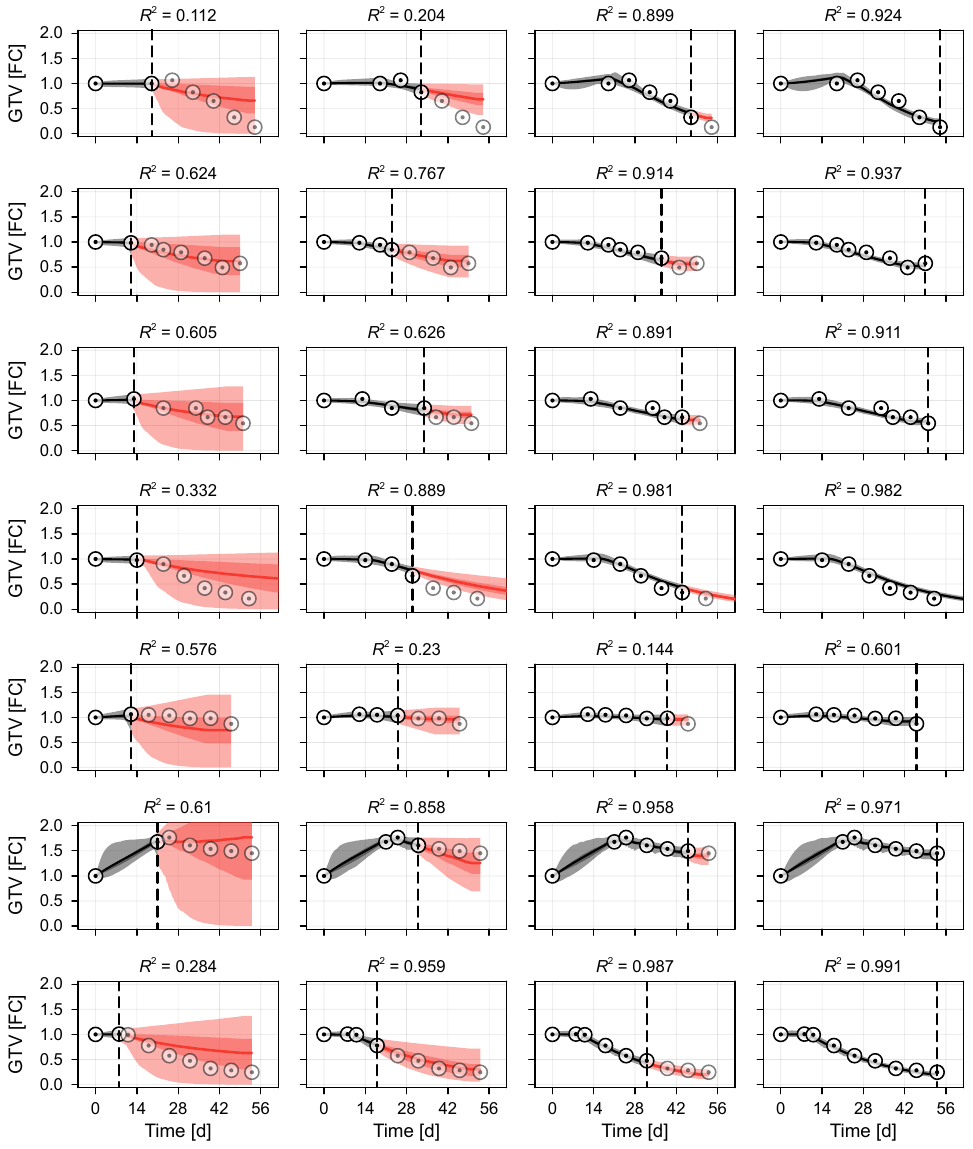}
		\caption[Figure S6]{ \textbf{Temporal predictions for the seven patients excluded from the training set that do not appear in Fig. 7 of the main document.} We reproduce the analysis from Fig. 5 and Fig. 7 in the main document for those remaining patients that were excluded from the training set. Each row corresponds to a single patient. These patients were not included in the training set, and so their results are representative of clinical predictions made throughout a new patient's course of treatment. }
		\label{figS6}
	\end{figure}
	
	\begin{figure}[h]
		\centering
		\includegraphics[width=\textwidth]{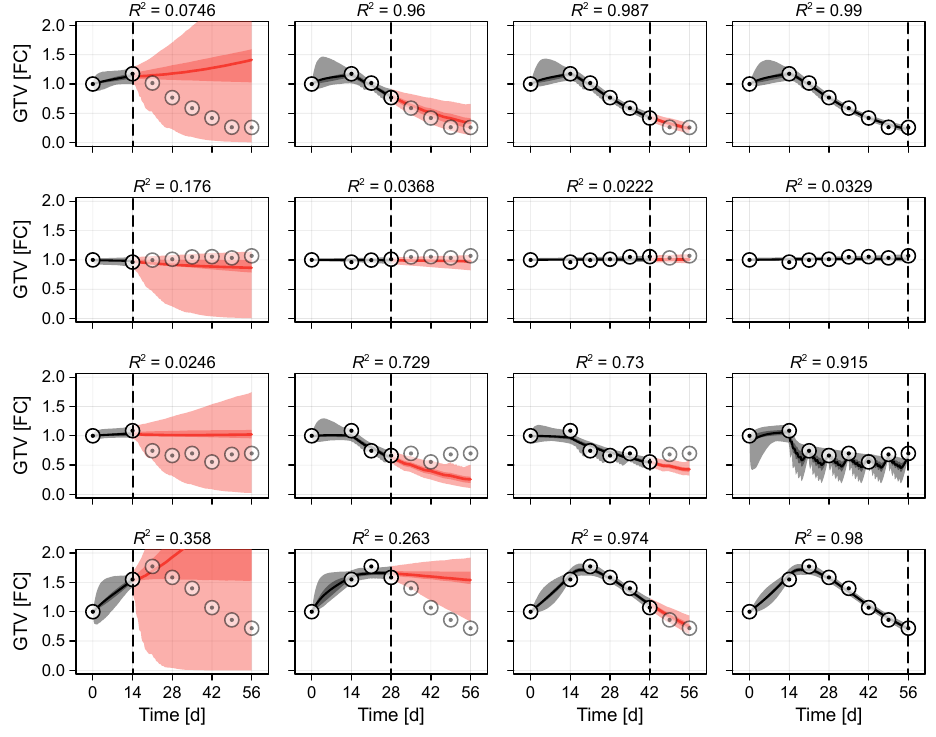}
		\caption[Figure S7]{ \textbf{Temporal predictions from four synthetic patients using the uninformative prior.} We reproduce the analysis from Fig. 5 in the main document using the uninformative (i.e., population-level) prior.}
		\label{figS7}
	\end{figure}

	\begin{figure}[h]
		\centering
		\includegraphics[width=\textwidth]{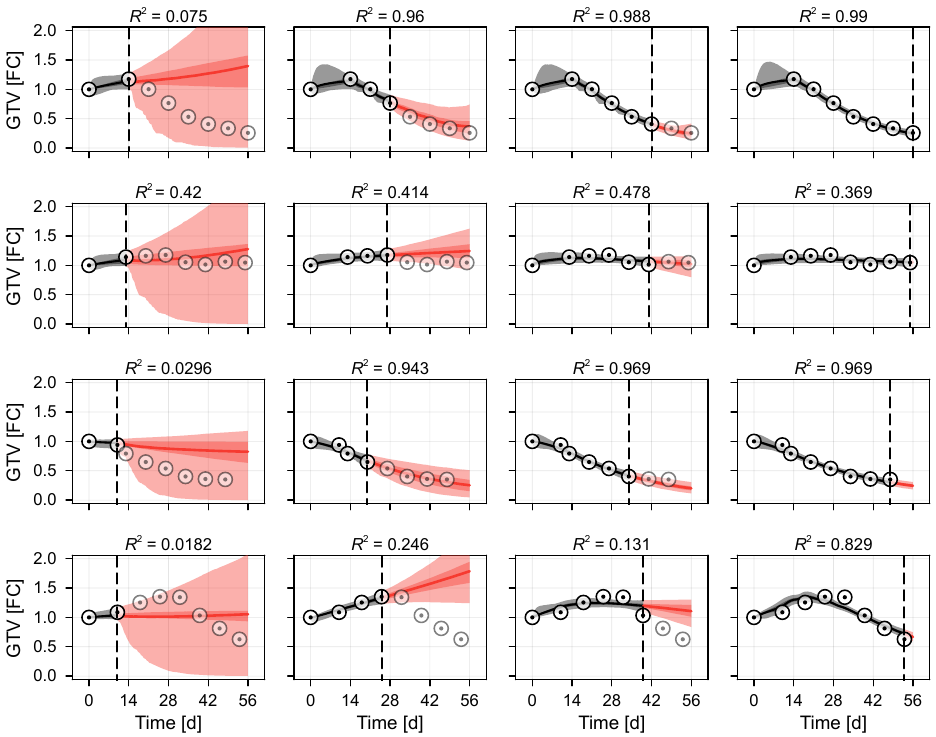}
		\caption[Figure S8]{ \textbf{Temporal predictions from four patients excluded from the training set using the uninformative prior.} We reproduce the analysis from Fig. 7 in the main document using the uninformative (i.e., population-level) prior.}
		\label{figS7}
	\end{figure}

\clearpage
\section*{Comparison to Bayesian hierarchical approach}
Here, we compare predicted uninformed patient trajectories from the pseudo-hierarchical approach presented in the main document, to a standard Bayesian hierarchical model implemented in \texttt{Turing.jl} \cite{ge2018t}. 

We assume that, at the population-level, the logarithm of each model parameter is distributed according to a truncated normal distribution, with support given by the support of the uniformed prior as implemented in the main text. For example,
	\begin{equation}\label{loglambdatrunc}
		\log \lambda \sim \mathrm{TruncatedNormal}(\mu_\lambda,\sigma_\lambda,\lambda_\text{min},\lambda_\text{max}),
	\end{equation}
where the prior for $\lambda$ in the main text was given by $\log \lambda \sim \mathrm{Uniform}(\lambda_\text{min},\lambda_\text{max})$. The population-level priors are given by
	\begin{subequations}
	\begin{align}
		\mu_\lambda &\sim \mathrm{Uniform}(\lambda_\text{min},\lambda_\text{max}),\\
		\log \sigma_\lambda &\sim \mathrm{Uniform}(-10.0,3.0),
	\end{align}
	\end{subequations}
where the prior for the log standard deviation is chosen to span a sufficiently wide range of scales that model parameters may either be concentrated, or approximatately uniformly distributioned (i.e., a truncated normal distribution with untruncated standard deviation much larger than the support of the truncated distribution). The priors for the other parameters $K$, $\gamma$, $\zeta$, $\eta$, and $\phi_0$ are given in the same way. The noise parameters, $\alpha_1$ and $\alpha_2$, are inferred simultaneously.

Since the individual dosing regime of each patient effectively prescribes a different mathematical model for each patient, the hierarchical problem is potentially much more computationally costly than the pseudo-hierarchical approach. As such, we demonstrate the hierarchical approach on a randomly chosen subset of the training data, comprising $\tilde{N} = 10$ patients. We perform inference using \texttt{Turing.jl}'s inbuilt Hamiltonian Monte Carlo algorithm (code available on GitHub\footnote{\url{https://github.com/ap-browning/clinical_predictions/blob/main/figS9.jl}}), and sample four independent chains, each of 50,000 samples. 

\begin{figure}[h]
		\centering
		\includegraphics{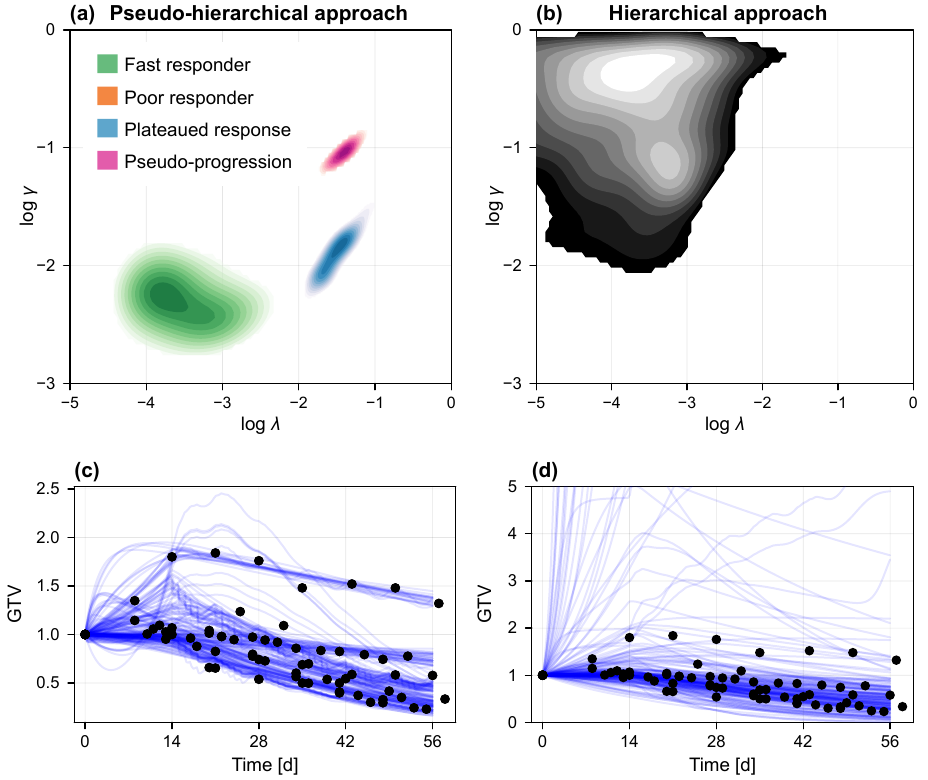}
		\caption[Figure S9]{\textbf{Comparison between the pseudo-hierarchical method and a standard Bayesian hierarchical method.} (a,b) Bivariate posterior distribution for patient-level parameters $\log \lambda$ and $\log \gamma$. In (a), parameter distributions are drawn from the mixture of individual posterior distributions for the 10 subjects in the resampled training set. In (b), the distribution is constructed by resampling from the posterior distributions for $\mu_\lambda$ and $\log \sigma_\lambda$, at each sample reconstructing and then sampling from the distribution for $\log \lambda$ given by \cref{loglambdatrunc} (and similar for parameters related to $\gamma$). (c,d) Predictions drawn from the patient-level distributions formed using (c) the pseudo-hierarchical approach, and (d) the hierarchical approach, for patients undergoing the standard course of treatment used to classify patients (see Section 2.2.1 of the main document). Also shown are GTV measurements from the $\tilde{N} = 10$ subjects in the resampled training set.}
		\label{figS9}
	\end{figure}

Results in \cref{figS9} highlight the fundamental differences between a standard hierarchical approach and the resampling-based approach presented in the main text. In particular, it is not straightforward to infer a correlation structure between the six patient-level parameters in a standard hierarchical approach, and so we do not gain information about parameter correlations or multimodality arising from the different patient responses (\cref{figS9}a,b). This is problematic for prediction, since the correlation structure and multimodality (recovered using the pseudo-hierarchical approach) constrains model predictions to the space of trajectories seen in the training data. We demonstrate this in3 \cref{figS9}c,d. Predictions from new patients (drawn before any patient-level data is observed) are often unrealistic, and are clearly not constrained to the set of observations drawn from the training data.

